\documentclass[a4paper,fleqn]{cas-sc}

\usepackage[numbers]{natbib}
\ExplSyntaxOn
\renewcommand{\printorcid}{
  \group_begin:
    \seq_if_empty:NF \g_stm_orcid_seq {
      \tex_let:D \thefootnote \relax \footnotetext {
        \raggedright
        \textsc{orcid}(s):\c_space_token
        \seq_use:Nn \g_stm_orcid_seq { ;~ }
      }
    }
  \group_end:
}
\ExplSyntaxOff

\usepackage{silence}
\usepackage{float} 
\usepackage{wrapfig}

\begin{document}
\let\WriteBookmarks\relax
\def\floatpagepagefraction{1}
\def\textpagefraction{.001}
\setlength{\emergencystretch}{3em}

\shorttitle{Complex Hyperspin Machine and Symmetry Quenching}

\shortauthors{Piacenti, Calvanese Strinati, and Conti}

\title[mode = title]{Complex Hyperspin Machine and Symmetry Quenching \\[0.5ex] \large Benchmark and scaling up to millions of degrees of freedom}

\affiliation[1]{organization={Physics Department, Sapienza University of Rome},
            addressline={Piazzale Aldo Moro 5},
            city={Rome},
            postcode={00185},
            country={Italy}}
\affiliation[2]{organization={Centro Ricerche Enrico Fermi (CREF)},
            addressline={Via Panisperna 89a},
            city={Rome},
            postcode={00184},
            country={Italy}}
            
\author[1]{Tommaso Piacenti}
\cormark[1]
\fnmark[1]
\ead{tommaso.piacenti@uniroma1.it}
\author[2]{Marcello {Calvanese Strinati}}
\fnmark[1]
\author[1]{Claudio Conti}

\cortext[1]{Corresponding author}
\fntext[1]{These authors contributed equally to this work.}
\begin{abstract}
Driven-dissipative oscillator networks have been proposed to simulate spin models on physical hardware, mostly focusing on systems of binary (Ising) spins, or systems of spin vectors in low dimensions. In this context, the equalized hyperspin machine was recently introduced to provide a single platform to simulate general models of $N$ coupled spin vectors with tunable dimension $D$, also realizing dynamical annealing protocols to vary the spin dimensionality in time. However, these schemes are limited to real spins coupled via real coupling matrices. Here, we propose a generalization of the hyperspin machine that works with complex fields and is able to deal with general complex Hermitian coupling matrices, also implementing an amplitude equalization scheme. This system, which we name complex hyperspin machine, is able to work on optimization problems with complex valued hyperspins that can be cast on complex $D$-vector Hamiltonians. We benchmark the machine against a differential evolution algorithm on complex $D=2$ hyperspins up to $N=80$, showing that the fraction of instances in which a lower-energy solution is found by the machine grows with system size and coupling density, reaching $92.5\%$ at the largest considered sizes. When the differential evolution finds a lower-energy solution, the machine still reaches a median relative energy deviation that lies between $10^{-15}$ and $10^{-12}$ while running at substantially lower computational cost. Furthermore, by exploiting the complex nature of the field, we introduce a new dimensional reduction protocol that we name symmetry quenching, which collapses complex variables that map to $D=2$ (XY) spins onto real $D=1$ (Ising) spins by switching on the parametric pump. We show on Max-Cut instances with known ground states that symmetry quenching outperforms a fixed-pump Ising machine, reaching exact ground states in the presence of equalization. Finally, we scale the system to $N=10^{6}$ hyperspins, or $8\times10^{6}$ dynamical variables, on a Hermitian sparse two-dimensional lattice.
\end{abstract}

\begin{highlights}
\item Hyperspin machine for optimization with complex Hermitian coupling
\item Symmetry quenching to enhance performance of Ising machines
\item Scaling up to one million multidimensional spins
\end{highlights}


{\hfuzz=120pt\maketitle}

\section{Introduction}\label{sec:Introduction}
Spin systems are of paramount importance in several areas of modern physics, ranging from fundamental studies to applications in combinatorial optimization. In the last years, finding efficient ways to find low-energy states of binary (Ising) spin models has attracted a renewed attention, mostly triggered by the possibility to map several combinatorial optimization problems onto Ising Hamiltonians~\cite{Lucas_2014_Ising_Formulation_of_Many_NP_Problems}: Finding the optimal solution with a huge number of interacting degrees of freedom translates into finding the ground state of a large network of interacting spins. These problems are notoriously hard~\cite{Barahona_1982}: The configuration space grows exponentially with the system size, and so does the computational time required to minimize the spin energy, which represents the cost function of the system. Because of this, complex combinatorial problems are often intractable on conventional computers even for moderate system size. This issue sparked the interest to find alternative computing methods to simulate spin models, leading to the conceptualization of Ising machines, namely, physical hardware designed to efficiently determine low-energy states (ideally, the ground state) of the simulated Ising spin system.

Specifically, Ising machines realized via coupled nonlinear oscillators have attracted significant attention in the last decade. These setups consist of networks of driven-dissipative degenerate parametric oscillators that can simulate systems of Ising spins by exploiting phase-dependent amplification: The phase of the amplified field admits two values only, which correspond to the two states ``up'' and ``down'' of an Ising spin (see Ref.~\cite{Strinati:26} for a review). In particular, the optical implementation of such systems, which relies on parametric down conversion to achieve phase-dependent amplification~\cite{Woo_1971_Fluctuations_in_a_Parametrically_Excited_Subharmonic_Oscillator,Marandi_2013_Coherent_Ising_Machine_Based_on_DOPOs}, currently represents a valuable platform to efficiently simulate Ising spin networks, operating at ultra-fast speed, room temperature and contained cost~\cite{Marandi_2013_Coherent_Ising_Machine_Based_on_DOPOs,Marandi_2014_Network_of_Time_Multiplexed_OPOs_as_CIM,Inagaki_2016_Large_Scale_Ising_Spin_Network_Based_on_OPOs,Yamamoto_2017_CIMs_Optical_NNs_Operating_at_the_Quantum_Limit,Yamamoto_2020_CIMs_Quantum_Optics_and_NN_Perspectives}. In recent years, oscillator-based systems to simulate coupled spin vectors have also been addressed. Most of the work has focused on the XY (planar) model: These XY machines rely on a phase independent amplification to encode in the continuous phase of the amplified fields the angle of the simulated XY spins. Notable implementations of XY machines involve non-degenerate optical parametric oscillators~\cite{Takeda_2018_Boltzsmann_Sampling_for_XY_Using_Non_Degenerate_OPOs_Network}, laser networks~\cite{Honarilatifpour_2020_Mapping_XY_Hamiltonian_onto_Network_of_Coupled_Lasers,Pal_2020_Rapid_Fair_Sapling_of_XY_Hamiltonian_with_Laser_Simulator,Gershenzon_2018_Exact_Mapping_Laser_Netwoek_Loss_Rate_and_Classical_XY_Hamiltonian_by_Laser_Loss_Control} and polariton systems~\cite{Berloff_2017_Realizing_XY_Hamiltonian_In_Polariton_Simulators}. 

The working principle of these machines is based on the fact that the dynamics of the nonlinear oscillators, when also coupled by a given matrix $\mathbb{J}$ which defines the network's graph, behaves as a gradient descent: Amplitudes and phases of the oscillators evolve in a way that minimizes a Lyapunov (cost) function $L$, which both includes single oscillator energies and their interactions. The dynamics is such that the interaction term of the Lyapunov function is proportional to the target spin Hamiltonian. Therefore, the time evolution drives the system towards an oscillator configuration that can be interpreted, under certain assumptions, as a low-energy state of the simulated spin model. This fact is precisely what allows us to use such networks in minimization tasks. 

In this framework the hyperspin machine was recently introduced in Refs.~\cite{CalvaneseStrinati_2022_Multidimensional_Hyperspin_Machine,CalvaneseStrinati_2024_Hyperscaling_in_the_Coherent_Hyperspin_Machine}. This system consists of a network of $ND$ oscillators, coupled both nonlinearly and linearly. The nonlinear coupling, arising for instance because of pump saturation where $D$ oscillators deplete one single shared pump~\cite{CalvaneseStrinati_2021_Can_NonLinear_POs_Solve_Random_Ising,Ben-Ami:23}, constrains the sum of the intensities of the $D$ fields to have a fixed value in the steady state of the dynamics. When this mechanism is applied to $N$ distinct sets of oscillators, the $ND$ oscillators are organized as $N$ vectors of $D$ components, where each oscillator field can be interpreted as a Cartesian component of a $D$-dimensional vector, called hyperspin. The linear coupling couples different $D$-dimensional spin vectors, inducing a scalar product with a selected metric. This scheme generalizes the paradigm of using oscillator systems as spin simulators, providing a single platform to simulate general vector spin models with an arbitrary number of components, with fully controllable coupling. The construction of the spin vectors via their individual components also allows the introduction of annealing protocols (e.g., dimensional annealing~\cite{CalvaneseStrinati_2022_Multidimensional_Hyperspin_Machine}) via direct control of the coupling components. 

A common problem for both the Ising and hyperspin machines based on oscillators is amplitude heterogeneity~\cite{Leleu_2019_Destabilization_of_Local_Minima_in_Analog_Spin_Systems_by_Correction_of_Amplitude_Heterogeneity}. Because of the fact that the oscillator amplitudes are in general different in the steady state, they do not allow the definition of spin vectors with equal amplitude. As a consequence, the system Lyapunov function does not perfectly map onto the target Hamiltonian~\cite{Roychowdhury_2022_Global_Lyapunov_for_CIM,CalvaneseStrinati_2021_Can_NonLinear_POs_Solve_Random_Ising,Leleu_2017_Combinatorial_Optimization_Using_Dynamical_Phase_Transitions_in_Driven_Dissipative_Systems,Leleu_2019_Destabilization_of_Local_Minima_in_Analog_Spin_Systems_by_Correction_of_Amplitude_Heterogeneity}.  
To correct this issue and achieve a proper mapping between system cost function and target Hamiltonian, amplitude equalization schemes have been proposed both for the Ising~\cite{Leleu_2019_Destabilization_of_Local_Minima_in_Analog_Spin_Systems_by_Correction_of_Amplitude_Heterogeneity} and the hyperspin machine~\cite{CalvaneseStrinati_2025_Equalized_Hyperspin_Machine}. These methods exploit an additional set of ``error correction variables'', or equalization fields, which are connected to the oscillator network and enforce via a dynamical feedback the desired amplitude constraint, in order to achieve in the steady state equal amplitudes for all oscillators in the case of the Ising machine, or equal hyperspin amplitudes in the case of the hyperspin machine. However, at the state of the art, such equalized machines have been proposed for real vector spin models only. The generalization of these systems for complex vectors is currently highly desirable, owing to their applicability in experimental contexts, e.g., phase retrieval~\cite{Wang_2026_Phase_Retrieval_via_Gain_Based_Photonic_XY_Hamilton_Optimization}, where complex coupling matrices naturally emerge.

In this paper, we introduce and validate the complex hyperspin machine. We make use of the inherent complex nature of the oscillator fields to generalize the equalized hyperspin machine proposed in Ref.~\cite{CalvaneseStrinati_2025_Equalized_Hyperspin_Machine}, which was built to only deal with real fields and real symmetric couplings, to a larger class of problems where the coupling matrix between the oscillators is in general complex. Moreover, complex fields carry twice the information compared to real fields. We show how this fact can be exploited to implement novel techniques for dimensionality tuning, which we name symmetry quenching, where the symmetry of the dynamical system is abruptly changed to effectively halve the dimensionality of the simulated spin vectors.

This paper is organized as follows. In Sec.~\ref{sec:Model}, we review the theoretical framework of the complex hyperspin machine with amplitude equalization. We highlight the main differences in the working principles of the machine with real couplings and complex couplings, and discuss how the machine can be used in different regimes depending on the task: with a non-zero pump the system variables are parametric oscillator fields and they can be used to deal with real coupling matrices either to reproduce the behavior of the equalized hyperspin machine~\cite{CalvaneseStrinati_2025_Equalized_Hyperspin_Machine}, or to implement symmetry quenching. With no parametric pump the system instead consists of coupled harmonic oscillators, and the machine can also be used to minimize the spin Hamiltonian with complex coupling matrices.

In Sec.~\ref{sec:Methods}, we describe our numerical methods. We discuss the parameter regimes of our simulations, and describe the set of complex Hermitian interaction graphs that we generate to benchmark the complex hyperspin machine. We focus on dense random graphs (RG) generated from a complex bimodal distribution with different densities, and on periodic square lattices (SL).

In Sec.~\ref{sec:Results}, we present our numerical results. We compare the machine's performance against a different minimizer that uses a differential evolution (DE) algorithm on systems of two-dimensional ($D=2$) complex hyperspins. The fraction of times the machine finds better energy minima than the DE algorithm is seen to grow with the overall problem size and complexity, while also offering a much faster performance. When the DE minimizer finds a lower energy state, we show that the machine consistently finds results that deviate from the DE minimum with a relative difference between $10^{-8}$ and $10^{-15}$ (maximum precision used in our simulations). We then introduce the symmetry quenching method: It exploits the fact that harmonic oscillators (i.e., parametric oscillators with no driving pump) can naturally be used to simulate a system of XY spins, which can then be collapsed to real binary variables (Ising spins) by turning on the parametric pump. The performance of symmetry quenching to solve Max-Cut problems, with coupling matrices from the BiqMac library~\cite{BiqMacLibrary}, is compared with that from the bare Ising machine simulation.

In Sec.~\ref{sec:Discussion} we present numerical data by scaling the system up to $N=10^6$ two-dimensional hyperspins on a sparse square lattice (in total $8\times 10^6$ degrees of freedom). We compare the performance as a minimizer of the $D$-vector model of the machine with and without equalization, showing that amplitude equalization provides a several orders of magnitude improvement of the determined minimum solution compared to the non-equalized system.

Conclusions are drawn in Sec.~\ref{sec:Conclusion}.

\section{Model}\label{sec:Model}

\subsection{Hyperspins by Parametric Oscillators}\label{sec:Hyperspins by Parametric Oscillators}

The interpretation of the steady state of an oscillator as a spin variable is here recalled for the sake of completeness. We specifically rely on Ref.~\cite{CalvaneseStrinati_2019_Theory_of_Coupled_POs} to derive a nonlinear equation that describes the behavior of an optical parametric oscillator (OPO). The equation for a classical oscillator field $x$ driven by a pump of amplitude $\tilde{h}$ and whose frequency is twice the proper oscillator frequency $\omega_0$ is
\begin{equation}\label{eqn:Single_Classical_PO}
\frac{d^2x}{dt^2}+\omega_0^2\left[1+\tilde{h}\sin(2\omega_0t)\right]x+\omega_0\gamma\left(g+\beta x^2\right)\frac{dx}{dt}=0 \,\, .
\end{equation}
The terms proportional to $g$ and $\beta$ represent the intrinsic loss of the medium and the cavity nonlinear coupling, respectively, and $\gamma$ is an arbitrarily small parameter. From Eq.~\eqref{eqn:Single_Classical_PO}, using the method of multiple scale analysis~\cite{Kevorkian_2012_Multiple_Scale_and_Singular_Perturbation_Methods}, we can isolate the slow part of the dynamics corresponding to the complex oscillator amplitude $A$, whose evolution is described by:
\begin{equation}\label{eqn:Single_Field_Non_Interacting_Equation}
\frac{d}{d t}A=\frac{h}{4}A^*-\frac{g}{2}A-\frac{\beta}{2}|A|^2A \,\, .
\end{equation}
The method consists in expanding the problem in terms of the small parameter $\gamma$. The parametric pump is rescaled in terms of $\gamma$ as $\tilde h=\gamma h$. One defines the slow time as $\tau=\gamma t$ and rewrites the time derivative:
\begin{equation}
    \frac{d}{dt}=\frac{\partial }{\partial t}+\gamma\frac{\partial}{\partial \tau}\,\, .
\end{equation}
The solution $x=x(t,\tau)$ is then expanded up to order $\gamma$ as
\begin{equation}
    x=x^{(0)}+\gamma x^{(1)}\,\, , 
\end{equation}
where $x^{(0)}$ is expressed in terms of the complex slowly varying amplitude $A=A(\tau)$ as
\begin{equation}
x^{(0)}(t,\tau)\equiv A(\tau)e^{i\omega_0t}+A^*(\tau)e^{-i\omega_0t}\,\, .
\end{equation}
The solvability condition for this problem, which requires that terms proportional to $e^{\pm i\omega_0t}$ vanish, yields Eq.~\eqref{eqn:Single_Field_Non_Interacting_Equation} by defining $\omega_0\tau\to t$. For a pump above the oscillation threshold value ($h>2g$), Eq.~\eqref{eqn:Single_Field_Non_Interacting_Equation} admits two steady state stable solutions $A_s$
\begin{equation}\label{eqn:Single_Field_Non_Interacting_Equation01}
A_s=\pm\sqrt{\frac{1}{\beta}\left(\frac{h}{2}-g\right)}\quad \,\,.
\end{equation}
The presence of two real solutions, symmetric with respect to the origin $A=0$, allows us to use the steady state of the oscillator to effectively encode a binary variable like an Ising spin, defined as $\sigma={\rm sign}(A)=\pm1$~\cite{CalvaneseStrinati_2019_Theory_of_Coupled_POs}.

This circumstance has been recently extended to realize multidimensional spin vectors (named hyperspin), by considering a set of $D$ complex fields $A_j$ that are coupled by an effective nonlinear coupling as~\cite{CalvaneseStrinati_2021_Can_NonLinear_POs_Solve_Random_Ising,CalvaneseStrinati_2022_Multidimensional_Hyperspin_Machine}
\begin{equation}\label{eqn:D_Fields_Equation}
\frac{d}{dt}A_j=\frac{h}{4}A_j^*-\frac{g}{2}A_j-\frac{\beta}{2}\left(\sum_{k=1}^{D}{\left|A_k\right|}^2\right)A_j\quad\quad\quad j=1,\dots,D \,\, .
\end{equation}
Such a nonlinear coupling can arise, for instance, when the oscillators saturate all the same pump~\cite{CalvaneseStrinati_2021_Can_NonLinear_POs_Solve_Random_Ising,Ben-Ami:23} or when they experience a nonlinear loss that involves all $D$ oscillators at once~\cite{CalvaneseStrinati_2024_Hyperscaling_in_the_Coherent_Hyperspin_Machine}. It can be convenient to introduce the vector $\vec{S}$, defined as:
\begin{equation}\label{eq:hyperspinvector01}
\vec{S}=(A_1,\dots,A_D) \,\, .
\end{equation}
Using $\vec{S}$ we can rewrite the dynamics in Eq.~\eqref{eqn:D_Fields_Equation} in a more compact way:
\begin{equation}\label{eqn:D_Parametric_Oscillators_Equation_Vector}
\frac{d}{dt}\vec{S}=\frac{h}{4}\vec{S}^*-\frac{g}{2}\vec{S}-\frac{\beta}{2}{\left|\!\left|\vec{S}\right|\!\right|}^2\vec{S} \,\, ,
\end{equation}
where $|\!|\vec{S}|\!|^2\equiv\sum_{k=1}^{D}|A_k|^2$ is the hyperspin squared norm. By equating Eq.~\eqref{eqn:D_Parametric_Oscillators_Equation_Vector} to zero, it follows that the stationary solutions of Eq.~\eqref{eqn:D_Parametric_Oscillators_Equation_Vector} are found on a $D$-dimensional hypersphere of radius whose value is given by $|A_s|$ in Eq.~\eqref{eqn:Single_Field_Non_Interacting_Equation01}. Each field appearing in $\vec{S}$ can be seen as one of the Cartesian components of a $D$-dimensional hyperspin (once normalized to have unitary norm). The hyperspin vector $\vec\sigma$ is then defined as:
\begin{equation}\label{eqn:Hyperspin_Variable_From_Fields}
\vec{\sigma}=\frac{\vec{S}}{\sqrt{{|\!|\vec S|\!|}^2}} \,\, .
\end{equation}
The hyperspin unit vector $\vec\sigma$ in Eq.~\eqref{eqn:Hyperspin_Variable_From_Fields} naturally lies on a hyperspherical surface of unit radius. This means that, when the oscillator fields are complex in the stationary state, we can interpret $\vec{\sigma}$ as a generally complex hyperspin. 

The hyperspin machine introduced in Ref.~\cite{CalvaneseStrinati_2022_Multidimensional_Hyperspin_Machine} simulates the dynamics of $N$, $D$-dimensional hyperspins by coupling several hyperspin vectors of the form in Eq.~\eqref{eq:hyperspinvector01} via a dissipative linear coupling as follows. Denoting by $A^{(p)}_j$ the $j$-th component of the $p$-hyperspin, we can write the system dynamics as:
\begin{equation}\label{eqn:Hyperspin_Machine_Fields_Dynamics}
\frac{d}{dt}A^{(p)}_j=\frac{h}{4}\left(A^{(p)}_j\right)^*-\frac{g}{2}A^{(p)}_j-\frac{\beta}{2}\left(\sum_{k=1}^{D}{\left|A^{(p)}_k\right|}^2\right)A^{(p)}_j+\frac{1}{2}\sum_{q=1}^{N}J_{pq}A_j^{(q)} \,\, ,
\end{equation}
where $J_{pq}$ are elements of the $N\times N$ coupling matrix $\mathbb{J}$. The literature on hyperspin machines has so far focused on real and symmetric $\mathbb{J}$ only. In this case, the oscillator fields in the steady state are real, and so are the hyperspin vectors in Eq.~\eqref{eq:hyperspinvector01}. For real $\mathbb{J}$, the system's dynamics can be used to minimize the corresponding $D$-vector model Hamiltonian of real valued spins~\cite{CalvaneseStrinati_2025_Equalized_Hyperspin_Machine}.

In this work, we extend the applicability of the hyperspin machine by considering the case in which $\mathbb{J}$ is a complex, Hermitian matrix ($J_{pq}=J^*_{qp}$) with $J_{qq}=0$ for all $q$ (no self-interaction). The oscillator fields in the steady state are in general complex, and thus the hyperspin vectors are in turn complex valued $D$-dimensional vectors. The cost function of the resulting network of complex valued hyperspins is then generalized as
\begin{equation}\label{eqn:D_Vector_Hamiltonian}
H_D=-\frac{1}{2}\sum_{p,q=1}^{N}\left[J_{pq}(\vec{\sigma_p}^*\cdot \vec{\sigma_q})+J_{pq}^*(\vec{\sigma_p}\cdot \vec{\sigma^*_q})\right] \,\, .
\end{equation}
The working principle of the complex hyperspin machine is that the dynamics described by Eq.~\eqref{eqn:Hyperspin_Machine_Fields_Dynamics}
leads to a configuration of fields that, once normalized to get hyperspin variables, corresponds to a low-energy value of Eq.~\eqref{eqn:D_Vector_Hamiltonian}. This follows from the fact that the system in Eq.~\eqref{eqn:Hyperspin_Machine_Fields_Dynamics} is a Lyapunov system, namely a scalar function $L$ can be defined such that it always decreases in time when computed along the trajectories of motion (see~\cite{Pukdeboon_2011_Lyapunov_Fundamentals}). For the complex hyperspin machine, we seek a Lyapunov function $L$ such that
\begin{equation}\label{eqn:Lyapunov_Condition}
\frac{\partial L}{\partial A^{(p)}_j}=-\frac{d (A^{(p)}_j)^*}{dt} \,\, .
\end{equation}
Using this definition, one sees that the time derivative of $L$ is always less or equal than zero (the function decreases over time):
\begin{equation}\label{eqn:Lyapunov_Decreasing}
\frac{dL}{dt}=\sum_{p=1}^{N}\sum_{j=1}^{D}\left(\frac{\partial L}{\partial A^{(p)}_j}\frac{d A^{(p)}_j}{dt}+\frac{\partial L}{\partial (A^{(p)}_j)^*}\frac{d (A^{(p)}_j)^*}{dt}\right)=- 2\sum_{p=1}^{N}\sum_{j=1}^{D} \left|\frac{d A^{(p)}_j}{dt}\right|^2\leq0  \,\, .
\end{equation}
In order to obtain the machine's equations of motion [Eq.~\eqref{eqn:Hyperspin_Machine_Fields_Dynamics}] as prescribed by condition in Eq.~\eqref{eqn:Lyapunov_Condition}, we define $L$ as:
\begin{equation}\label{eqn:Lyapunov_Function}
L=L_0+L_{\text{int}}\,\, ,
\end{equation}
where:
\begin{equation}\label{eqn:Lyapunov_Definitions}
\begin{split}
&L_0=-\frac{1}{2}\sum_{q=1}^{N}\left[\frac{h}{4}\sum_{k=1}^{D}\left[(A^{(q)}_k)^2+(A^{(q)*}_k)^2\right]-g{\left|\!\left|\vec{S_q}\right|\!\right|}^2-\frac{\beta}{2}{\left|\!\left|\vec{S_q}\right|\!\right|}^4\right]\,\, ,\\
&L_{\text{int}}=-\frac{1}{4}\sum_{p,q=1}^N\left[J_{pq}(\vec{S_p}^*\cdot \vec{S_q})+J_{pq}^*(\vec{S_p}\cdot \vec{S^*_q})\right] \,\, .
\end{split}
\end{equation}
When in the stationary state the amplitudes of the hyperspin vectors $\vec S_p=(A^{(p)}_1,\dots,A^{(p)}_D)$ are all equal, i.e., $|\!|\vec{S}_p|\!|\equiv S$ for all $p$, the term $L_{\rm int}$ in Eq.~\eqref{eqn:Lyapunov_Definitions} becomes proportional to $H_D$ in Eq.~\eqref{eqn:D_Vector_Hamiltonian}. The condition in Eq.~\eqref{eqn:Lyapunov_Decreasing} implies that the system dynamics minimizes $H_D$ in Eq.~\eqref{eqn:D_Vector_Hamiltonian}, because the Lyapunov function $L$ in the equal amplitude case reads $L\equiv L_0+S^2H_D/2$, where $L_0$ is the non-interacting part of the Lyapunov function in Eq.~\eqref{eqn:Lyapunov_Definitions} computed for equal hyperspin amplitudes.

\subsection{Amplitude Heterogeneity and Equalization}\label{sec:Amplitude Heterogeneity and Equalization}
When the norms of different hyperspin fields are not equal, the system exhibits amplitude heterogeneity~\cite{Leleu_2019_Destabilization_of_Local_Minima_in_Analog_Spin_Systems_by_Correction_of_Amplitude_Heterogeneity}. This causes the system dynamics to minimize a Lyapunov function $L$ that is not properly mapped onto $H_D$, therefore hampering the system's ability to minimize the target Hamiltonian.

For the real hyperspin machine, Ref.~\cite{CalvaneseStrinati_2025_Equalized_Hyperspin_Machine} proposes a method to equalize the hyperspin amplitudes, by coupling the oscillator fields to additional variables, called equalizers, that are coupled to the oscillators via a non-linear and antisymmetric coupling. We here generalize the equalization mechanism to the complex hyperspin machine. For clarity, we first consider the simple case of two interacting $D$-dimensional hyperspins ($N=2$), coupled to two complex scalar equalization fields $Y_1$ and $Y_2$. With two hyperspins only, amplitude heterogeneity is induced, e.g., by pumping the hyperspins with two different pump fields $h_1$ and $h_2$~\cite{CalvaneseStrinati_2025_Equalized_Hyperspin_Machine}. The modified equations for the dynamics of the hyperspin fields, coupled to the equalizers, are:
\begin{equation}\label{eqn:Equalized_Two_Hyperspin_Fields_Vector_Equation}
\begin{cases}
\displaystyle
\frac{d}{dt}\vec{S_A}=\frac{h_1}{4}\vec{S_A}^*-\frac{g}{2}\vec{S_A}-\frac{\beta}{2}{\left|\!\left|\vec{S_A}\right|\!\right|}^2\vec{S_A}+\frac{J}{2}\vec{S_B}-\frac{\delta_{eq}}{2}\left({|Y_1|}^2-{|Y_2|}^2\right)\vec{S_A}
\\\\
\displaystyle
\frac{d}{dt}\vec{S_B}=\frac{h_2}{4}\vec{S_B}^*-\frac{g}{2}\vec{S_B}-\frac{\beta}{2}{\left|\!\left|\vec{S_B}\right|\!\right|}^2\vec{S_B}+\frac{J^*}{2}\vec{S_A}+\frac{\delta_{eq}}{2}\left(|Y_1|^2-|Y_2|^2\right)\vec{S_B} \,\, ,
\end{cases}
\end{equation}
where $\delta_{eq}$ is the equalization coupling constant. Furthermore, the equations for the dynamics of the equalization fields are given by:
\begin{equation}\label{eqn:Equalizators_Two_Dynamics}
\begin{cases}
\displaystyle
\frac{dY_1}{dt}=\frac{\delta_{eq}}{2}\left({\left|\!\left|\vec{S}_A\right|\!\right|}^2-{\left|\!\left|\vec{S}_B\right|\!\right|}^2\right)Y_1\\
\\
\displaystyle
\frac{dY_2}{dt}=-\frac{\delta_{eq}}{2}\left({\left|\!\left|\vec{S}_A\right|\!\right|}^2-{\left|\!\left|\vec{S}_B\right|\!\right|}^2\right)Y_2 \,\, .
\end{cases}
\end{equation}
In Eqs.~\eqref{eqn:Equalized_Two_Hyperspin_Fields_Vector_Equation} and~\eqref{eqn:Equalizators_Two_Dynamics}, equalization of the hyperspin amplitudes occurs as a consequence of the antisymmetric nature of the coupling between oscillator and equalizer fields. 
This coupling is such that as soon as the hyperspin field amplitudes differ, one equalizer is amplified and the other is suppressed with a rate that is proportional to the hyperspin amplitude difference. This introduces in the hyperspin fields equations a feedback that reduces the larger one and amplifies the smaller, pushing the system towards an equalized steady state~\cite{CalvaneseStrinati_2025_Equalized_Hyperspin_Machine}.
Moreover, one can see from Eq.~\eqref{eqn:Equalizators_Two_Dynamics} that the stationary state condition imposes $|\!|\vec{S}_A|\!|^2=|\!|\vec{S}_B|\!|^2\equiv S^2$, i.e., a steady state with equal hyperspin amplitudes. 

\begin{figure} 
    \centering
    \includegraphics[width=16.4cm]{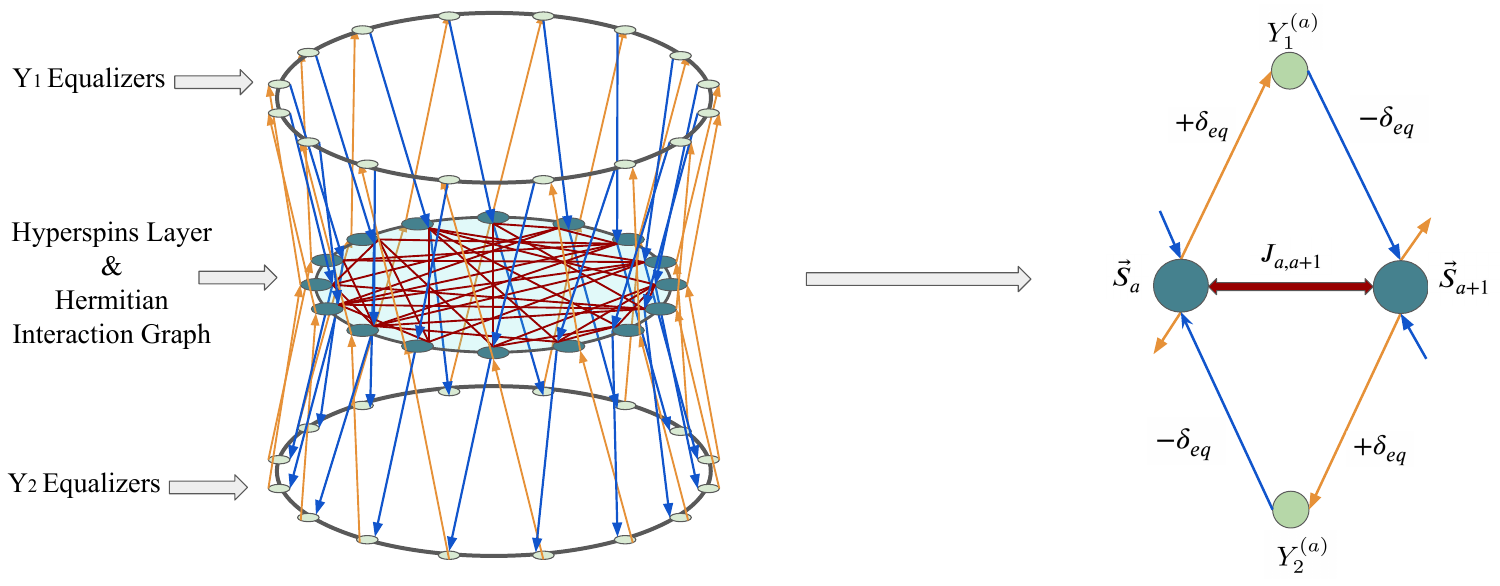}
    \caption{Schematic representation of the network that composes the complex equalized hyperspin machine and detail of a single equalization pair acting on two interacting hyperspins. (Left) Composite system of hyperspins and equalizers, including the hyperspin fields that interact according to an arbitrary Hermitian graph (central layer), and the two equalization rings (closed loops above and below the hyperspin layer) for the $Y^{(a)}_1$ and $Y^{(a)}_2$, respectively. (Right) Details of the equalization couplings, which are real and directional (specifically antisymmetric), where $\delta_{\rm eq}$ defines the strength of the coupling between hyperspins and equalizers. The different sign appearing in front of $\delta_{\rm eq}$ is shown and depicted by arrows of different colors.}
    \label{fig:Equalization_Chain}
\end{figure}

In the general case of $N$, $D$-dimensional complex hyperspins, we consider $2N$ equalization fields, coupled to the oscillators over a closed equalization loop, as shown in Fig.~\ref{fig:Equalization_Chain} (at variance with Ref.~\cite{CalvaneseStrinati_2025_Equalized_Hyperspin_Machine} where equalizers were organized on an open loop). The equalizers are organized as $N$ pairs indexed by $a$, so that the $a$-th hyperspin interacts with the $a$-th and the $(a-1)$-th equalizers (it is intended that $a=0$ coincides with $a=N$ due to periodic boundary conditions). For a given pair, we use the indices ``1'' and ``2'' to identify the two equalization fields ($Y_1^{(a)},Y_2^{(a)}$). The equation for a generic oscillator field $A^{(a)}_j$ is then given by
\begin{equation}\label{Machine_Hyperspin_Field_General_Equation}  
\begin{split}
\frac{d A^{(a)}_j}{dt}=&\frac{h}{4}{\left(A^{(a)}_j\right)}^*-\frac{g}{2}A^{(a)}_j-\frac{\beta}{2}\left(\sum_{k=1}^{D}{\left|A^{(a)}_k\right|}^2\right)A^{(a)}_j+\frac{1}{2}\sum_{b=1}^{N}J_{ab}A^{(b)}_j+ \\
+&\frac{\delta_{eq}}{2}\left[\left({\left|Y_1^{(a)}\right|}^2-{\left|Y_2^{(a)}\right|}^2\right)-\left({\left|Y_1^{(a-1)}\right|}^2-{\left|Y_2^{(a-1)}\right|}^2\right)\right]A^{(a)}_j \,\, .
\end{split}
\end{equation}
Moreover, the equation for a generic couple of equalizers is
\begin{equation}\label{Machine_Equalization_Fields_General_Equation}  
\begin{cases}
\displaystyle
\frac{d Y^{(a)}_1}{dt}=\frac{\delta_{eq}}{2}\left[{\left|\!\left|\vec{S}_{a+1}\right|\!\right|}^2-{\left|\!\left|\vec{S}_a\right|\!\right|}^2\right]Y^{(a)}_1\\
\\
\displaystyle
\frac{d Y^{(a)}_2}{dt}=-\frac{\delta_{eq}}{2}\left[{\left|\!\left|\vec{S}_{a+1}\right|\!\right|}^2-{\left|\!\left|\vec{S}_a\right|\!\right|}^2\right]Y^{(a)}_2 \,\, .
\end{cases}
\end{equation}
The structure of the oscillator network that composes the complex hyperspin machine is shown in Fig.~\ref{fig:Equalization_Chain}.

\subsection{Complex and real hyperspins}\label{sec:Complex Hyperspins}
Before discussing our numerical data, it is crucial to discuss a key difference between real and complex hyperspin machine. The oscillator fields in Eq.~\eqref{eqn:Hyperspin_Machine_Fields_Dynamics} are in general complex. However, when all system quantities (pump, loss term, nonlinearity, and coupling matrix) are real and the pump amplitude is non-zero, the dynamics suppresses the imaginary part of the oscillators, effectively reducing to real oscillator fields~\cite{CalvaneseStrinati_2019_Theory_of_Coupled_POs}, and therefore real hyperspin vectors~\cite{CalvaneseStrinati_2022_Multidimensional_Hyperspin_Machine}. Instead, when considering $\mathbb{J}$ complex, the coupling mixes real and imaginary parts of the oscillators. As such, the full complex dynamics of the system has to be taken into account. In particular, because of the fact that the pump term is proportional to the complex conjugate field on the right-hand side of Eq.~\eqref{eqn:Hyperspin_Machine_Fields_Dynamics} (phase-dependent amplification), we need to discuss the way the parametric pump influences the behavior of the complex hyperspin machine.

To proceed, we consider the dynamics in Eq.~\eqref{eqn:Hyperspin_Machine_Fields_Dynamics} in the linear regime of operation (near-threshold regime), and label the oscillator fields by the index $j=1,\ldots,DN$ as
\begin{equation}\label{eqn:Hyperspin_Machine_Fields_Dynamics_General_Complex_Pump}
\frac{d A_j}{d t}=\frac{h}{4}A_j^*-\frac{g}{2}A_j+\frac{1}{2}\sum_{k=1}^{ND}\mathbb{K}_{jk}A_k  \,\, ,
\end{equation}
where $\mathbb{K}=\mathbb{J}\otimes\mathbb{I}_D$ is the $DN\times DN$ coupling matrix (here $\mathbb{I}_D$ is the $D\times D$ identity matrix and ``$\otimes$'' denotes the Kronecker product). By stacking all real and imaginary parts of the oscillator amplitudes $A_j=A_{R,j}+iA_{I,j}$ as column vectors $\vec{A}_R$ and $\vec{A}_I$, respectively, and by separating real and imaginary parts of the coupling matrix $\mathbb{K}=\mathbb{K}^R+i\mathbb{K}^I$, we can rewrite the dynamics of the linear system as
\begin{equation}\label{eqn:Linearized_Dynamics}
\frac{d\mathbf{A}}{dt}=\mathbb{S}\,\mathbf{A} \quad\mbox{with}\quad \mathbf{A}=\begin{pmatrix}
\vec{A_R}\\\vec{A_I}\end{pmatrix}\,\, ,
\end{equation}    
where $\mathbb{S}$ is the $(2ND)\times(2ND)$ matrix given by
\begin{equation}\label{eqn:Linearized_Dynamics_Matrix}
    \mathbb{S}\equiv\frac{h}{4}\,\hat z\otimes\mathbb{I}_{ND}-\frac{g}{2}\,\mathbb{I}_2\otimes\mathbb{I}_{ND}+\frac{1}{2}\,\mathbb{I}_2\otimes\mathbb{K}^R-\frac{i}{2}\,\hat y\otimes\mathbb{K}^I \,\, .
\end{equation}
In Eq.~\eqref{eqn:Linearized_Dynamics_Matrix} we denote the Pauli matrices as
\begin{equation}
\hat x=\begin{pmatrix}
0 & 1 \\ 1 & 0\end{pmatrix}\qquad 
\hat y=\begin{pmatrix}
0 & -i \\i & 0\end{pmatrix}\qquad
\hat z=\begin{pmatrix}
1 & 0 \\0 & -1\end{pmatrix}\qquad
\mathbb{I}_2=\begin{pmatrix}
1 & 0 \\0 & 1\end{pmatrix} \,\, .
\end{equation}
Notice that since $\mathbb{J}$ is Hermitian (so $\mathbb{K}$ is Hermitian), then $\mathbb{K}^R$ is symmetric and $\mathbb{K}^I$ is skew-symmetric. This means that in general $\mathbb{K}^R$ and $\mathbb{K}^I$ do not commute, and that the linearized matrix $\mathbb{S}$ is symmetric. The oscillation threshold is defined as the smallest value of $h$, denoted as $h_{\rm th}$, such that the largest eigenvalue of $\mathbb{S}$ turns from negative to positive, and accordingly the oscillator amplitudes, starting from very small values, are amplified in time following Eq.~\eqref{eqn:Linearized_Dynamics}. Close to threshold, the system dynamics steers the oscillator amplitudes to the configuration dictated by the eigenvector of $\mathbb{S}$ associated with the largest eigenvalue (also dubbed the maximal eigenvector $\mathbf{A}_{\rm eig}$). One can see that $\mathbb{S}\equiv \mathbb{S}(h)$ does not in general commute with $\mathbb{S}(h=0)$, which is in turn proportional to the coupling matrix. This is because both Pauli matrices $\hat{y}$ and $\hat{z}$ appear in $\mathbb{S}$ for nonzero $h$ and $\mathbb{K}^I$, while for $h=0$ the Pauli matrix $\hat{z}$ is absent. Therefore, in general, the steady state of the system in the near-threshold regime is not trivially related to the eigenvectors of the coupling matrix. This phenomenology differs from that of the Ising~\cite{Leleu_2017_Combinatorial_Optimization_Using_Dynamical_Phase_Transitions_in_Driven_Dissipative_Systems,CalvaneseStrinati_2021_Can_NonLinear_POs_Solve_Random_Ising} or hyperspin machines~\cite{CalvaneseStrinati_2022_Multidimensional_Hyperspin_Machine} close to threshold, which are conventionally studied for real coupling only, where the system dynamics converges towards the maximal eigenvector of $\mathbb{K}$. In these cases, it has been reported that such a maximal eigenvector defines a spin configuration via Eq.~\eqref{eqn:Hyperspin_Variable_From_Fields} that is a low-energy state (albeit not the ground state in general) of the simulated spin model.

Instead, in the complex hyperspin machine, we observed that, when $h>h_{\rm th}$ and the coupling matrix is complex (both $\mathbb{K}^R$ and $\mathbb{K}^I$ are nonzero), the hyperspin state $\vec{\sigma}_{\rm eig}$ defined from $\mathbf{A}_{\rm eig}$ using Eqs.~\eqref{eq:hyperspinvector01} and~\eqref{eqn:Hyperspin_Variable_From_Fields} has an energy [Eq.~\eqref{eqn:D_Vector_Hamiltonian}] that is significantly above the global minimum, suggesting that the dynamics for nonzero pump is not guaranteed to resemble the one of the corresponding hyperspin system. For the purpose of the present work, we can overcome this issue by setting $h$ to zero when simulating complex coupling matrices. In this case, the system turns into a network of coupled harmonic oscillators. From Eq.~\eqref{eqn:Hyperspin_Machine_Fields_Dynamics_General_Complex_Pump}, we see that the effective gain is provided by the coupling, i.e., the dynamics is amplified if $\lambda_{\rm max}>g$, where $\lambda_{\rm max}$ is the largest eigenvalue of $\mathbb{K}$. In the following, the value of $g$ is such that the system is above threshold (amplified) also for zero pump. Setting $h=0$ allows us to simulate the complex hyperspin machine above threshold while avoiding the Pauli matrix $\hat{z}$ to affect the dynamics: The matrix $\mathbb{S}$ is then proportional to the coupling matrix, sharing with it an eigenbasis, and in analogy with the case of real systems, close to threshold, the complex hyperspin machine relaxes towards a low-energy state of the complex $D$-vector model in Eq.~\eqref{eqn:D_Vector_Hamiltonian}.

\section{Methods}\label{sec:Methods}
\subsection{Details on the numerical integration}\label{sec:Runge-Kutta integration}
To simulate the complex hyperspin machine with equalization in Eqs.~\eqref{Machine_Hyperspin_Field_General_Equation} and~\eqref{Machine_Equalization_Fields_General_Equation} we develop a C-language software development kit (publicly available at \url{https://github.com/tommasopiacenti/Complex_Hyperspin_Machine}). The dynamics is integrated using a second-order Runge-Kutta algorithm for a sufficiently long time $T_{\rm max}$, with time step $dt=\{0.01, 0.02,0.05,0.1\}$ depending on the case, to ensure that the system dynamics reaches a stationary state. The real and imaginary parts of the oscillator fields are initialized as random numbers from a uniform distribution on $[-A_{\max},A_{\max}]$, with $A_{\max}=0.1$ or $A_{\max}=1.0$. The nonlinear coefficients $\beta$ and $\delta_{eq}$ are chosen in the range $[0,1]$. The entries of the coupling matrices are always normalized so that each element is of the order $1/n_{int}$, where $n_{int}$ denotes the average number of interacting neighbors that each hyperspin has (the degree of the graph). Both the pump amplitude and the loss coefficient are chosen in the range $h\in[0,1]$ and $g\in[0,1]$.

Numerical simulations are implemented using a sparse format for coupling matrices. This significantly speeds up the numerical integration, the computation of observables, and optimizes the memory usage: The number of operations in an integration step, as well as the number of stored couplings, is reduced from order of $N^2$ to $Nn_{int}$, allowing us to simulate remarkably large system sizes up to $10^6$ hyperspins, as detailed below, especially when the graph matrix $\mathbb{J}$ is very sparse ($n_{int}$ is sufficiently smaller than $N$). Simulations are run on several processors by resorting to the \texttt{OpenMP} library. In our simulations, we stop the Runge-Kutta integration at time $T_{\rm max}$ when the energy variation
\begin{equation}\label{eqn:Energy_Difference}
\Delta H(t)\equiv\left|{H(t+dt)-H(t)}\right| \,\, ,
\end{equation}
takes values smaller than a given numerical tolerance (between $10^{-12}$ and $10^{-14}$, close to machine precision) for a sufficient number of consecutive checks (usually $100$ to $1000$ checks, where a check is performed every $10$ time steps).

\subsection{Graph Families}\label{sec:Graph Families}
In order to assess the efficiency of the complex hyperspin machine on Hermitian coupling matrices, we generate our own set of benchmark graphs from two main families: (i) Random graphs (RGs) and (ii) Periodic square lattices (SLs) of linear size $L$ such that $N=L^2$. Both families are defined by a parameter $\alpha$: It quantifies the average fraction of couplings that have a  positive real part over the total number of existing non-zero entries of $\mathbb{J}$. The parameter $\alpha$  is used to extract the sign of the real part of the couplings ($\alpha\in[0,1]$) and the imaginary parts are determined from it as we will explain shortly. The value of $\alpha$ can be interpreted as a quantification of the frustration level of a given graph. For a real matrix we indeed have that $\alpha=1$ corresponds to a purely ferromagnetic interaction (hyperspin alignment along the same direction is preferred), while with lower values of $\alpha$ we start having some level of frustration (some couplings become antiferromagnetic).

\begin{figure} 
    \centering
    \includegraphics[width=16.4cm]{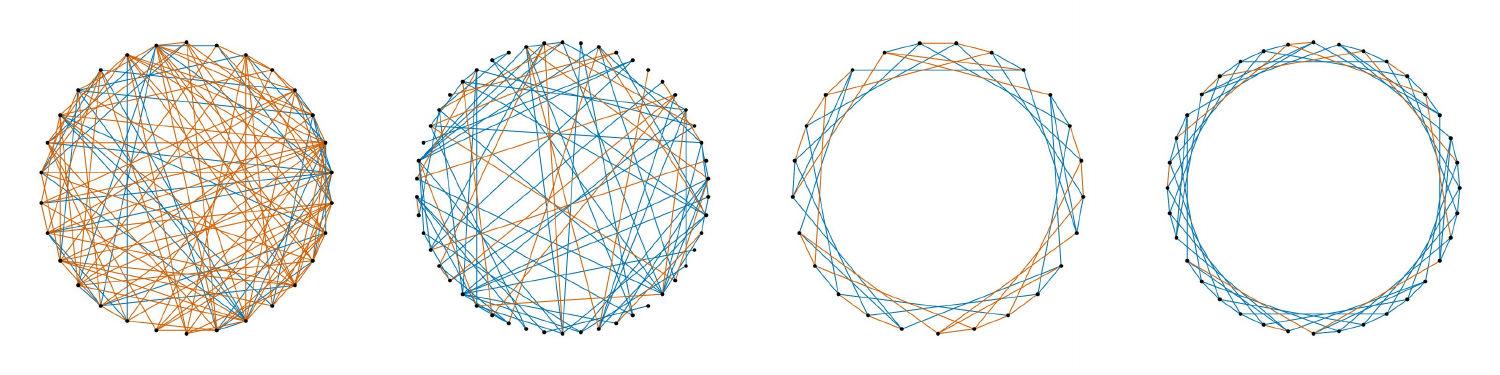}
    \caption{Representation of the Hermitian interaction graphs in the complex bimodal family, where the graph nodes are arranged on a circular embedding. Blue lines represent ferromagnetic interactions, orange lines represent antiferromagnetic interactions. The two leftmost graphs are random graphs generated with $N=30$, $\rho=0.3$, $\alpha=0.25$, and $N=50$, $\rho=0.1$, $\alpha=0.75$, respectively. The rightmost ones represent square lattices with a fixed connectivity (each site has precisely 4 neighbors), generated with $N=25$, $\alpha=0.5$, and $N=36$, $\alpha=0.75$, respectively.}
    \label{fig:Example_Graphs}
\end{figure}

Figure~\ref{fig:Example_Graphs} shows examples of the representations of these graphs. In an SL graph with periodic boundary conditions, nodes are arranged on a torus by setting as nonzero only the elements of the coupling matrix corresponding to the nearest neighbor connections in each direction (so every node has degree equal to four, i.e., it is connected to four other nodes). Instead, the RG graphs are defined by an additional parameter, i.e., the connectivity $\rho$, which represents the average edge probability ($\rho\in(0,1]$). Specifically, if $\rho=1$ the graph is fully connected, while if $\rho\sim1/N$ or less we deal with a sparse one. For both the RGs and SLs graphs, we normalize the couplings according to
\begin{equation}\label{eqn:Couplings_Normalization}
\begin{cases}
\displaystyle
J_{pq}^{RG}=\frac{z}{\rho N\sqrt{2}}\\
\\
\displaystyle
J_{pq}^{SL}=\frac{z}{4\sqrt{2}}
\end{cases} \,\, .
\end{equation}
In Eq.~\eqref{eqn:Couplings_Normalization}, $z$ is a random complex number extracted for each $p$ and $q$ from a complex bimodal distribution. Considering only the upper triangle of the coupling matrix, we have two possible values for $z$: $1-i$ and $-1+i$.

\section{Results}\label{sec:Results}

In this section we present the main results we obtained by the complex hyperspin machine.

\subsection{Complex Hyperspins Benchmarks with Hermitian Coupling Matrices}\label{sec:Complex Hyperspins Tests with Hermitian Coupling Matrices}
In order to assess the machine's capability to find good energy minima for the complex $D$-vector Hamiltonian in Eq.~\eqref{eqn:D_Vector_Hamiltonian}, we compare the hyperspin energy at the end of the dynamics with the minimum value of $H_D$ found from another numerical minimizer. We here focus on complex hyperspins of dimension $D=2$, so in total each hyperspin vector consists of four real degrees of freedom. For each coupling matrix $\mathbb{J}$, the minimum energy value of Eq.~\eqref{eqn:D_Vector_Hamiltonian} (which we call $H_{\rm min}$) that we use to benchmark the machine is found by a differential evolution algorithm (DE), which is here specifically implemented by the numerical minimizer \texttt{NMinimize[]} by Wolfram Mathematica~\cite{Mathematica}.

Similar to what was reported in other numerical contexts with the real hyperspin machine~\cite{CalvaneseStrinati_2024_Hyperscaling_in_the_Coherent_Hyperspin_Machine,CalvaneseStrinati_2025_Equalized_Hyperspin_Machine}, we observe that the complex hyperspin machine is able to reach lower energy values than the numerical minimizer, for fixed minimization parameters (accuracy, precision, machine working precision,  number of DE search points). The occurrence with which this happens depends both on the graph family and on the system variables (complex or real hyperspins). When generating our benchmark graphs, we observed that the fraction of matrices for which the complex hyperspin machine finds a lower energy state than the DE algorithm ranges from $0$ to $92.5\%$ depending on the family, the size and the density, as shown in Table~\ref{tab:Minimizer_Failure_Fraction}.

\begin{table} 
    \centering
    \begin{tabular}{|c||c|c|c|c|c|}
\hline
N & $\rho=0.1$ (\%) & $\rho=0.3$ (\%) & $\rho=0.5$ (\%) & $\rho=0.7$ (\%) & $\rho=0.9$ (\%)\\ 
\hline
10 & 0 & 0 & 0 & 0 & 0 \\
20 & 0 & 0 & 0 & 0 & 0 \\
30 & 0 & 2.5 & 0 & 0 & 0 \\
40 & 0 & 2.5 & 0 & 7.5 & 0 \\
50 & 0 & 0 & 5.0 & 40.0 & 67.5 \\
60 & 7.5 & 5.0 & 40.0 & 65.0 & 70.0 \\
70 & 2.5 & 17.5 & 60.0 & 72.5 & 85.0 \\
80 & 0 & 60.0 & 85.0 & 92.5 & 92.5 \\
 \hline
    \end{tabular}
    \caption{Fraction (in percentage) of the matrices for which the complex hyperspin machine finds a lower energy state than the DE numerical minimizer (see text). These matrices are excluded from the benchmark because at least one of the machine runs violates the condition in Eq.~\eqref{eq:relativetolerance}. The simulations are all performed using the same set of parameters both for the machine and the minimizer. We consider a total of $40$ matrices per size $N$. For each $N$, the dynamics of the complex hyperspin machine is simulated $30$ times. These data are extracted from the results corresponding to the Hermitian random graph family, with $\alpha=0.75$. These data showcase the ability of the complex hyperspin machine to perform well in minimization tasks.} 
    \label{tab:Minimizer_Failure_Fraction}
\end{table}

\begin{figure} 
    \centering
    \includegraphics[width=16cm]{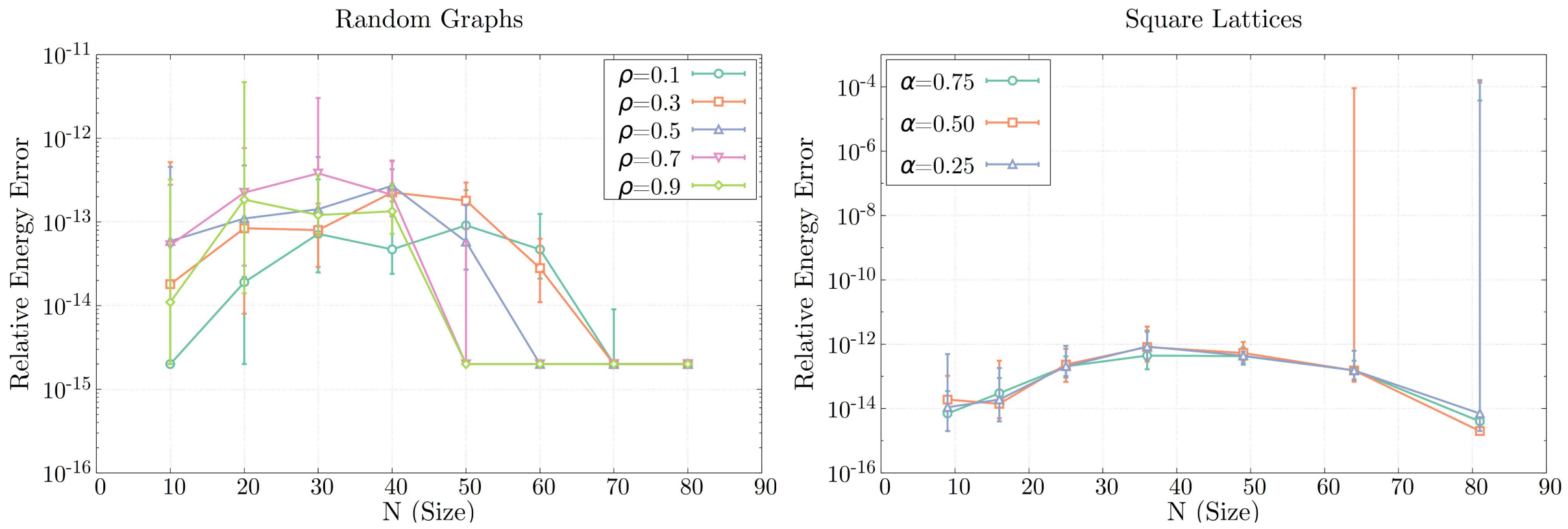}
    \caption{Relative error in Eq.~\eqref{eqn:Relative_Error} as function of the system size $N$, for coupling instances where the hyperspin machine finds higher energy values than $H_{\rm min}$ by the DE algorithm (see text). The complex hyperspin machine matches the best energy $H_{\rm min}$, with a relative error typically within the range $[10^{-15},10^{-12}]$ for systems of $D=2$ complex hyperspins. The coupling matrices are generated according to the complex bimodal distribution, with probability of having a positive real part of the coupling denoted by $\alpha$. We consider two graph geometries: (Left) Random graphs with $\alpha=0.75$ and density $\rho=0.1,0.3,0.5,0.7,0.9$ up to $N=80$, and (Right) Periodic square lattices with $\alpha=0.25,0.50,0.75$ up to $N=81$. Points show the median over several matrices and hyperspin machine runs per matrix, and vertical bars represent the inter-quartile range (IQR). Matrix instances in which the machine outperforms the DE algorithm in finding the lowest energy were excluded if at least one machine run violated the condition in Eq.~\eqref{eq:relativetolerance}. Consistently with this minimizer acceptance condition, $2\times10^{-15}$ was used as floor value for these plots. Consequently, the number of contributing matrices varies with family, $N$ and $\rho$. The flattening of the RG results in the left panel at large sizes near the minimum allowed \text{RelErr} value is primarily a consequence of the condition imposed on the minimizer. This demonstrates that for these graph types, the machine either outperforms the DE algorithm or matches its results. The large interquartile range (IQR) observed in the case of square lattices in the right panel is ascribed to the fact that the machine reaches high-energy metastable states during its dynamics.}
    \label{fig:RelErr_vs_N_Hermit_RG_pm075_and_LT}
\end{figure}

The remarkable outcome from Table~\ref{tab:Minimizer_Failure_Fraction} is that the complex hyperspin machine in a significant number of cases provides a more reliable minimizer. In the remaining cases, where the hyperspin machine converges to an energy value above or equal to $H_{\min}$, we benchmark the machine as follows: We discard data for which at least one machine run violates the following tolerance condition (recall that $H_{\rm min}<0$ in general)
\begin{equation}\label{eq:relativetolerance}
    \frac{H-H_\text{min}}{H_\text{min}}\leq2\times10^{-15} \,\, .
\end{equation}
Furthermore, we quantify the machine performance by the relative energy difference (RelErr)
\begin{equation}\label{eqn:Relative_Error}
    \text{RelErr}=\left|\frac{H-H_\text{min}}{H_\text{min}}\right| \,\, .
\end{equation}
Figure~\ref{fig:RelErr_vs_N_Hermit_RG_pm075_and_LT} shows the median relative energy difference as function of $N$ for the random graph family up to size $N=80$ for $\alpha=0.75$ and for the square lattice up to size $N=81$ (i.e., $L=3,\ldots,9$) for $\alpha=0.25,0.50,0.75$. Scaling to even larger sizes was not possible with our numerical resources because of the increasing computational cost required by the DE minimizer. We find that the median relative energy difference takes values between $10^{-15}$ and $10^{-12}$ for all the graph instances generated according to the complex bimodal distribution. For these benchmarks the hyperspin dynamics was simulated using $dt=0.01$, $A_{\max}=0.1$, $\beta=0.1$, $g=0$ and $\delta_{eq}=1.0$. The energy tolerance $\Delta H(t)$ defined in Eq.~\eqref{eqn:Energy_Difference} was set to $10^{-13}$, with the requirement to be satisfied for $10^{3}$ consecutive integration steps. The large IQR values for the lattice sizes $N=64$ and $N=81$ on the right panel are ascribed to a particularly rugged energy landscape: The machine often finds the correct minimum, but sometimes the dynamics is trapped in higher energy relative minima. 

\begin{figure} 
    \centering
    \includegraphics[width=16.4cm]{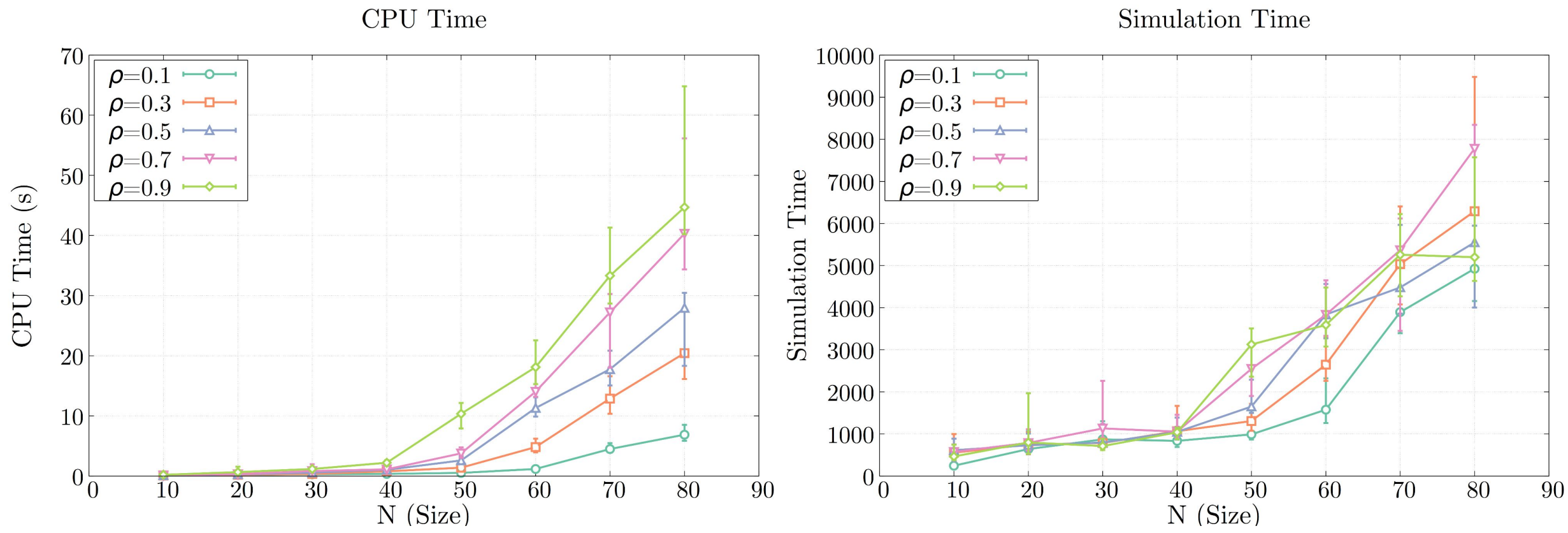}
    \caption{(Left) CPU time in seconds and (Right) Maximum integration time $T_{\rm max}$ in dimensionless units (see Sec.~\ref{sec:Runge-Kutta integration}) required for convergence as a function of system size $N$ for the Hermitian random-graph family with $\alpha=0.75$. A run is considered converged when $\Delta H(t)$ in Eq.~\eqref{eqn:Energy_Difference} remains less than $10^{-13}$ for $10^{3}$ consecutive steps. Curves correspond to edge probabilities $\rho=0.1,0.3,0.5,0.7,0.9$ for system sizes up to $N=80$. At each size, points represent the median of the per-matrix mean times, and vertical bars represent the inter-quartile range (IQR). The median CPU time remains below one minute at all sizes and edge probabilities taken into consideration. CPU times are measured using an ``11th Gen Intel(R) Core(TM) i9-11900K'' processor (nominal 3.50 GHz) running on a single thread. Matrix instances in which the machine outperformed the DE algorithm in finding the lowest energy were excluded. Consequently, the number of contributing matrices varies with $N$ and $\rho$.}
    \label{fig:Times_vs_N_Hermit_RG_CPU_and_Sim.png}
\end{figure}

Figure~\ref{fig:Times_vs_N_Hermit_RG_CPU_and_Sim.png} shows the median CPU time as function of the system size, as well as the maximum integration time $T_{\rm max}$ discussed in Sec.~\ref{sec:Runge-Kutta integration}. Simulations are performed using an ``11th Gen Intel(R) Core(TM) i9-11900K'' processor (nominal 3.50 GHz) operating without multi-threading (no parallelization). The CPU time is therefore measured as the actual computational time, in seconds, taken by our hardware to finish a given simulation. Considering the fact that the energy determined by the hyperspin machine often coincides with $H_\text{min}$, up to machine precision, we run $30$ simulations in order to be sure to reach the same minimum on these families. We observe that, while the minimization with the DE algorithm could take up to an hour for $N=80$ systems (not shown), the hyperspin machine is able to reach similar energy values in minutes. For the simulated values of $N$, the simulation of the complex hyperspin machine takes a fraction of the time required by the DE algorithm.

These results show that, at least for these custom generated complex bimodal graph families, the complex equalized hyperspin machine proves to be a competitive tool for optimization, capable of dealing in relatively short times with problems involving a large number of degrees of freedom. The case of $N=80$ and $D=2$ indeed corresponds to a total of $320$ dynamical variables counting both real and imaginary components of the fields, plus the additional $320$ degrees of freedom that come from the complex equalization fields.

We remark that the capability of the complex hyperspin machine to find lower energy values compared to the DE numerical minimizer, together with the fact that the dynamics usually converges to a stationary state faster than the minimization algorithm, hints that the machine provides a rather stable and reliable method to determine good energy states for optimization problems that can be cast in the form of a $D$-vector Hamiltonian in Eq.~\eqref{eqn:D_Vector_Hamiltonian}, at least for this complex bimodal family. This allows us to find good solutions for problems of increasing sizes that may not be easily accessible with the other numerical minimization methods.

\subsection{Symmetry Quenching Method and Max-Cut Problems}\label{sec:Symmetry Quenching Methods for Max-Cut Problems}
We now validate the complex hyperspin machine as a solver of the Ising model with real spins. It has been shown that the hyperspin machine can significantly enhance the accuracy and precision of the oscillator system to converge towards the ground state of the simulated Ising model compared to the Ising machine by employing an effective spin dimensionality reduction, named dimensional annealing~\cite{CalvaneseStrinati_2022_Multidimensional_Hyperspin_Machine,CalvaneseStrinati_2024_Hyperscaling_in_the_Coherent_Hyperspin_Machine}. The rationale is the following: Since the ground state of the target Ising Hamiltonian ($D=1$) with a given $\mathbb{J}$ is in general an excited state of a vector spin system with the same $\mathbb{J}$, one can first simulate spins in a higher dimensional space ($D>1$) exploiting the additional dimensions to escape local energy minima of the target Ising model. Then the spin dimensionality is reduced during the dynamics by gradually zeroing selected hyperspin components corresponding to a specific Cartesian direction (see Ref.~\cite{Strinati:26} for a review). In Refs.~\cite{CalvaneseStrinati_2022_Multidimensional_Hyperspin_Machine,CalvaneseStrinati_2024_Hyperscaling_in_the_Coherent_Hyperspin_Machine}, this dimensionality reduction has been shown to boost the performance of the system as an Ising machine, specifically interpolating from $D=2$ (real XY spins) to $D=1$ (real Ising spins) by reducing to zero selected components of the full coupling matrix $\mathbb{K}$ [see Eq.~\eqref{eqn:Hyperspin_Machine_Fields_Dynamics_General_Complex_Pump}].

In this section, we take inspiration from this argument and introduce a new method for dimensional reduction, which we name symmetry quenching. At variance with previous proposals of dimensional annealing, which are based on the effective dimensionality reduction of real spin vectors by acting on the coupling, symmetry quenching exploits the presence of the complex conjugate field in Eq.~\eqref{Machine_Hyperspin_Field_General_Equation}, detailing phase-dependent amplification, to effectively project spins on a lower-dimensional space by changing the symmetry of the equations of motion. For $h=0$, the system has continuous time-translational symmetry: Each complex oscillator field $A_j$ describes a harmonic oscillator, whose phase $\varphi_j$ can take in general any value between $0$ and $2\pi$ [$U(1)$ symmetry due to phase-independent amplification]. An oscillator field can then be expressed as a real, two-dimensional XY vector with $x$- and $y$-Cartesian components given by
\begin{equation}\label{eq:xytoising01}
A_j=|A_j|e^{i\varphi_j} \quad\rightarrow\quad \vec{A}_j=|A_j|\left(\begin{array}{c}\cos(\varphi_j)\\\sin(\varphi_j)\end{array}\right) \,\, .
\end{equation}
For $h>0$, the presence of a parametric pump induces discrete time-translational symmetry in the model~\cite{CalvaneseStrinati_2019_Theory_of_Coupled_POs}: The $U(1)$ symmetry abruptly reduces to $\mathbb{Z}_2$ as the pump is turned to a nonzero value, and the phase of each oscillator takes only two values, $\varphi_j=0,\pi$. The $y$-component of the two-dimensional vector in Eq.~\eqref{eq:xytoising01} describing the imaginary part of the field vanishes, so that the field is real and the $x$-component of the vector in Eq.~\eqref{eq:xytoising01} encodes a real Ising spin (defined along $x$).

\begin{figure} 
    \centering
    \includegraphics[width=16.4cm]{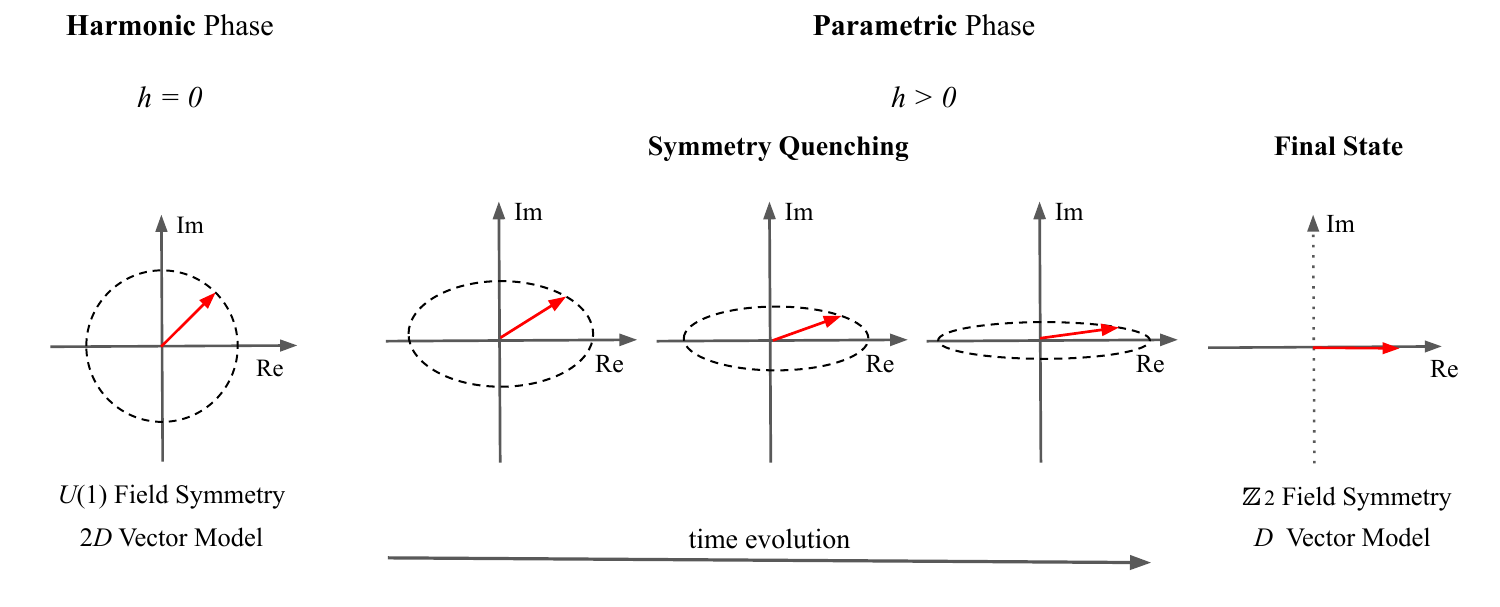}
    \caption{Schematic visualization of the two phases of the complex hyperspin machine with real symmetric couplings. The machine initially operates in the harmonic phase, when the parametric pump is off. In this case the real and imaginary components of all hyperspin fields are equally amplified and can converge to any point on a circle in the complex plane. The system's variables enjoy a U(1) phase symmetry. As soon as the pump is nonzero ($h\neq 0 $) the system operates in the parametric phase ($\mathbb{Z}_2$ symmetry): the pump amplifies the oscillators' real parts and damps the imaginary ones, leading to a final state that lies on the real axis with only two allowed phases $0$ and $\pi$. The symmetry quenching can be performed both with and without equalization. The quenching scheme shown in Fig.~\ref{fig:Trajectory_Plots} uses a linear increasing pump.}
    \label{fig:CHSM_Quenching}
\end{figure}

This interpretation extends to the case of $ND$ complex fields as follows. First, the dynamics expressed in terms of $N$, $D$-dimensional complex vectors $\vec{S}_j$ from Eq.~\eqref{Machine_Hyperspin_Field_General_Equation} can be recast by stacking for each complex hyperspin $\vec{S}_j=\vec{S}^R_j+i\vec{S}^I_j$ the real and imaginary parts ($\vec{S}^R_j$ and $\vec{S}^I_j$, respectively) into one real $(2D)$-dimensional vector as
\begin{equation}\label{eq:xytoising02}
\vec{V}_j=\left(\begin{array}{c}\vec{S}^R_j\\\vec{S}^I_j\end{array}\right) \,\, ,
\end{equation}
for each $j$. Second, we see that in Eq.~\eqref{Machine_Hyperspin_Field_General_Equation} for real $\mathbb{J}$, the real and imaginary parts of different oscillator fields are not linearly coupled. Therefore, the complex hyperspin machine of $N$, $D$-dimensional complex vectors $\vec{S}_j$ coupled via a real $\mathbb{J}$ naturally remaps onto a real hyperspin machine of $N$, $(2D)$-dimensional real vectors $\vec{V}_j$ in Eq.~\eqref{eq:xytoising02} coupled by the same $\mathbb{J}$. When $h=0$, the system operates in the harmonic phase, where $\vec{V}_j$ expresses a collection of $ND$ harmonic oscillators: All components of the $(2D)$-dimensional vectors in Eq.~\eqref{eq:xytoising02} are in general nonzero. When $h>0$, the system enters the parametric phase: The imaginary parts of the hyperspin vectors vanish, so that the vectors in Eq.~\eqref{eq:xytoising02} are effectively projected onto a $D$-dimensional real vector. This mechanism therefore gives rise to a dimensional reduction of the hyperspin space, from $2D$ to $D$ (see Fig.~\ref{fig:CHSM_Quenching}).

To test the complex hyperspin machine performance against known problems, we perform symmetry quenching from $D=1$ complex to $D=1$ real hyperspins, which according to Eq.~\eqref{eq:xytoising02} corresponds to projecting real vectors with two components (XY spins) onto vectors with one component (Ising spins). This allows us to simulate the Ising model starting from XY spins. The machine accuracy is then quantified as the probability with which the system converges at the end of the quenching on the ground state of the Ising model simulated by a given $\mathbb{J}$.

We proceed with this analysis by taking benchmark matrices for the Ising model from the BiqMac library~\cite{BiqMacLibrary,BiqMacLibraryReport,BiqMacSolver}, which includes a large collection of solved optimization problems. We recall that these problems are expressed as cut maximization (Max-Cut) problems, and their relation to the Ising minimization problems is here reviewed for the sake of completeness. Max-Cut problems are usually expressed in terms of a weight matrix $\omega_{ij}=0,1$, where the quantity to maximize (the cut), is expressed as:
\begin{equation}
    C(\vec \sigma) = \frac{1}{2}\sum_{i<j}\omega_{ij}(1-\sigma_i\sigma_j)\equiv M-\frac{1}{2}\sum_{i<j}\omega_{ij}\sigma_i\sigma_j \,\, ,
\end{equation}
where $M\equiv\frac{1}{2}\sum_{i<j}\omega_{ij}$ and $\sigma_j=\pm1$ (real Ising spin). According to our Hamiltonian definition in Eq.~\eqref{eqn:D_Vector_Hamiltonian}, for a real symmetric $\mathbb{J}$, and considering real Ising spins ($D=1$), we have $\omega_{ij}=-4J_{ij}$. This way we can express $H(\vec \sigma)=M-C(\vec \sigma)$, where $H(\vec{\sigma})$ is the Ising Hamiltonian for a specific Ising spin configuration $\vec{\sigma}$. Therefore, cut maximization and Ising Hamiltonian minimization are related as
\begin{equation}\label{eqn:Max-Cut_to_Ising}
H_{\rm min}=M-C_{\rm max} \,\, .
\end{equation}
The $\omega_{ij}$ weights from the Max-Cut library and the corresponding optimal values were rescaled to match our normalization scheme described in Eq.~\eqref{eqn:Couplings_Normalization}.

\begin{figure} 
    \centering
    \includegraphics[width=16.4cm]{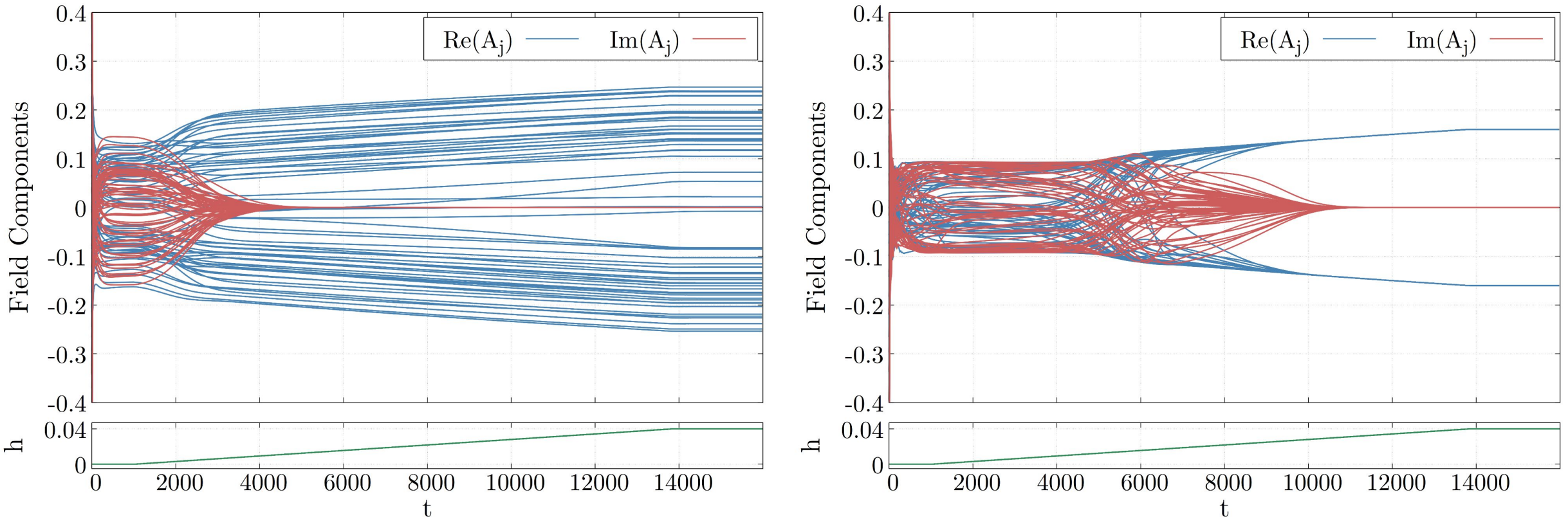}
    \caption{(Top panels) Real (blue) and imaginary (red) components of $N=60$, $D=1$ hyperspin fields as function of time from the simulation of Eqs.~\eqref{Machine_Hyperspin_Field_General_Equation} and~\eqref{Machine_Equalization_Fields_General_Equation}. (Bottom panels) The parametric pump $h$ during the dynamics, which ramps from zero to $h_\text{max}=0.04$. Top left panel shows the evolution in absence of equalization ($\delta_{eq}=0$), while in the top right panel equalization is performed with $\delta_{eq}=1$. The dynamics is simulated on a graph corresponding to the Max-Cut $\mathrm{g}05\_60.0$ instance~\cite{BiqMacLibraryReport}. The machine is run using a time step $dt=0.02$, a loss coefficient $g=0.05$, a nonlinear coupling $\beta=0.9$ and random complex initial conditions for the fields. The pump is kept constant at $h=0$ from $t=0$ to $t=1000$, then every $\Delta t=80$ time units the pump is increased by $\Delta h =2.5\times 10^{-4}$. At $h=0$ (harmonic phase) the system is invariant under a $U(1)$ rotation of the complex fields as both the real and imaginary components are equally amplified by the coupling. In these conditions the system variables behave as XY spins with no preferred phase. As soon as the parametric pump becomes nonzero ($h>0$) the system changes from $U(1)$ (XY spins) to $\mathbb{Z}_2$ (Ising spins) as the imaginary parts of the fields are attenuated. The collapse to Ising spins is faster and happens earlier in the non-equalized case in the left panel. This is ascribed to the fact that equalization introduces a non-linear driving force that keeps the imaginary components relevant for longer times as the pump ramps.}
    \label{fig:Trajectory_Plots}
\end{figure}

Figure~\ref{fig:Trajectory_Plots} depicts the real and imaginary components of the fields reacting to the dynamical change of the parametric pump from zero to $h_\text{max}$. During the time evolution of the equalized complex hyperspin machine, the pump value $h$ can be dynamically switched on and off during a run to transition between harmonic and parametric phases. Each complex hyperspin can freely explore the phase space when the parametric pump is initially set to zero (harmonic phase). As soon as the pump is nonzero, the system symmetry reduces to $\mathbb{Z}_2$ (parametric phase). In the tests we performed, we use $g$ such that the system is amplified also for zero pump, and the value of $h$ starts from zero and then is slowly increased until a fixed value. We observe that as soon as the pump ramp begins the imaginary components survive longer when equalization is turned on. This can be explained by the fact that the coupling between oscillators and equalizers acts as a nonlinear drive that opposes the parametric damping. This phenomenon however only affects the transitioning time between the harmonic and the $\mathbb{Z}_2$ symmetry phases. This is because with any nonzero $\beta$ the nonlinear saturation effectively couples real and imaginary components: The real parts of the fields are amplified by the pump, and the imaginary parts are suppressed.

Symmetry quenching allows the machine to reach rather satisfying results in finding the Ising energy minimum on graphs from the BiqMac Max-Cut library. Table~\ref{tab:BiqMac_Rudy_Results} shows the success probability (SP) of the complex hyperspin machine using the symmetry quenching method for all the matrices in the ``$\mathrm{g05\_60}$'' set (see Table~6 of Ref.~\cite{BiqMacLibraryReport}). We here define the SP as the fraction of runs that reach an energy with a relative difference from the known ground state of at most $10^{-6}$. During each run, three quenching cycles, that we denote by $c_1$, $c_2$, and $c_3$, are performed as follows:
\begin{enumerate}
    \item The system evolves with $h=0$ from $t=0$ to $t=1000$;
    \item The pump ramp starts, linearly increasing $h$ from zero to $h_\text{max}=0.04$ in a span of $12800$ time units. The energy is read at the end of the ramp. The pump is then abruptly reset to zero;
    \item After $c_1$, step 2 is repeated two more times for $c_2$ and $c_3$. At the end of $c_3$ the pump is kept constant at $h=h_\text{max}$;
    \item The simulation then stops when an energy equilibrium stop condition is met, specifically we require the energy difference in Eq.~\eqref{eqn:Energy_Difference} to be less than $10^{-10}$ for $1000$ consecutive integration steps.
\end{enumerate}

\begin{table}[t]
    \centering
    \begin{tabular}{|c||c|c|c|c|}
\hline
$\mathrm{g05}$ & $c_1$ (\%) & $c_2$ (\%) & $c_3$ (\%) & any (\%) \\
\hline
 60.0 & 10.0 & 6.7 & 3.3 & 10.0 \\
 60.1 & 63.3 & 76.7 & 90.0 & 90.0 \\
 60.2 & 63.3 & 83.3 & 93.3 & 93.3 \\
 60.3 & 53.3 & 76.7 & 76.7 & 76.7 \\
 60.4 & 40.0 & 36.7 & 46.7 & 73.3 \\
 60.5 & 60.0 & 33.3 & 80.0 & 96.7 \\
 60.6 & 0.0 & 16.7 & 0.0 & 16.7 \\
 60.7 & 80.0 & 96.7 & 100.0 & 100.0 \\
 60.8 & 6.7 & 6.7 & 13.3 & 26.7 \\
 60.9 & 3.3 & 6.7 & 0.0 & 10.0 \\
 \hline
    \end{tabular}
    \caption{Success probability (SP) of the complex hyperspin machine on the ten ``$g05\_60$'' instances of the Max-Cut set~\cite{BiqMacLibraryReport} ($N=60$, $D=1$, real antiferromagnetic couplings), with $30$ independent machine runs per instance. The runs use the symmetry quenching protocol with equalization at $\delta_{eq}=0.2$, a time step $dt=0.02$, a loss coefficient $g=0.05$ and a nonlinear coefficient $\beta=0.9$, starting from general complex initial conditions. Each of the three quenching cycles ramps the pump from $h=0$ to $h_\text{max}=0.04$. The $c_1$, $c_2$ and $c_3$ columns report the SP at the energy readings taken immediately before each of the two pump resets ($t=13799$ and $t=26599$), and at the end of the last cycle. A run is considered successful when the energy found by the machine lies within a relative deviation $10^{-6}$ of the known ground-state energy $H_\text{min}$ for that instance. The last column gives the fraction of runs that reach $H_\text{min}$ at least once across the three readings.}
    \label{tab:BiqMac_Rudy_Results}
\end{table}
\begin{figure}
    \centering
    \includegraphics[width=12cm]{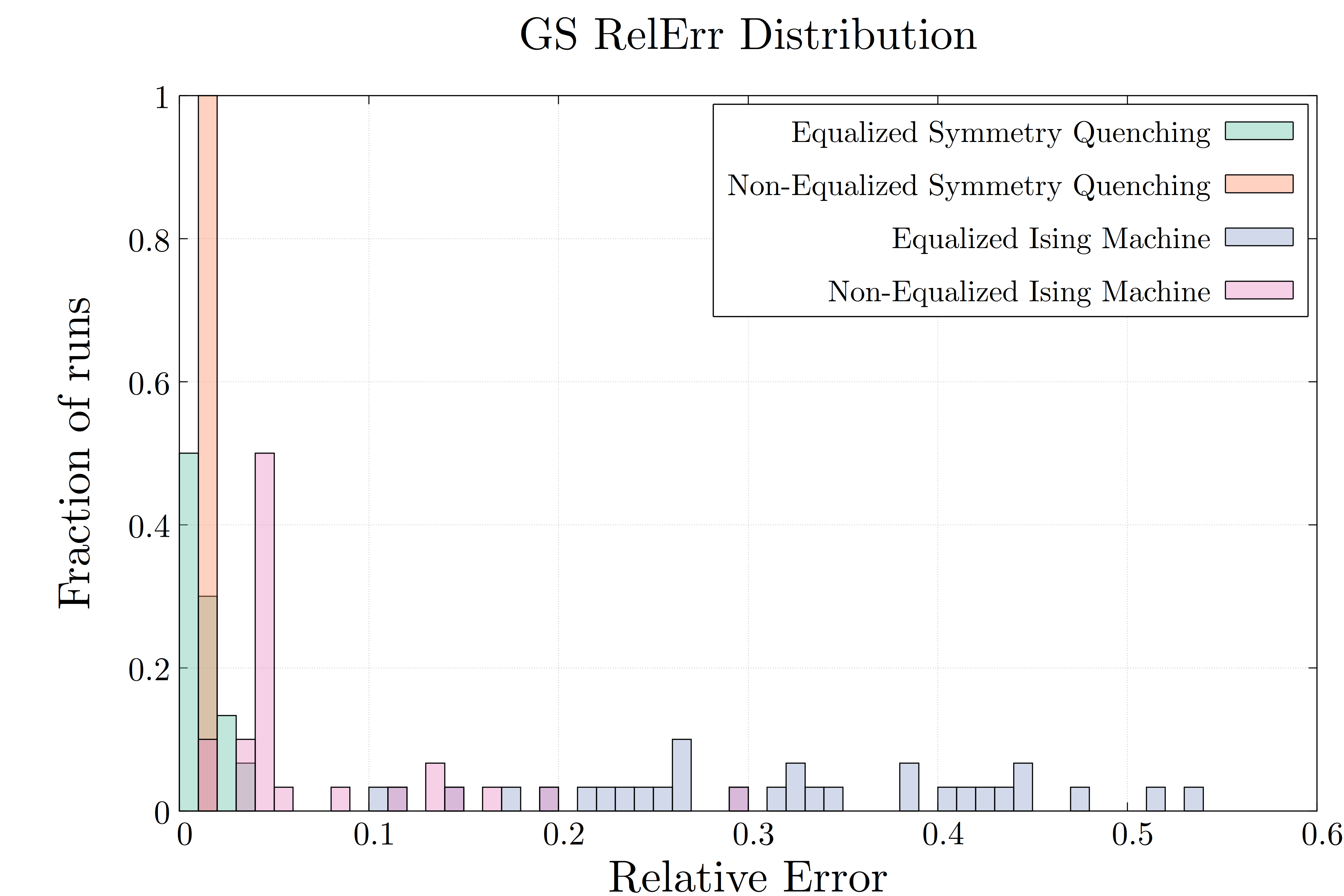}
    \caption{Histogram of the relative error of the machine energy from the known ground state for the $\mathrm{g}05\_60.0$ graph instance~\cite{BiqMacLibraryReport}. Results for four setups of the complex hyperspin machine are shown with $30$ runs each: (i) Equalized and non-equalized symmetry quenching from XY to Ising spins (green and orange bars, respectively, using $dt=0.02$, $g=0.05$, varying $h$ from $0$ to $h_\text{max}=0.04$), and (ii) Equalized and non-equalized Ising machine (blue and pink bars, with $dt=0.01$, $g=0.5$, $h=0.8$). In the Ising machine case, the imaginary parts of the fields are quickly damped: The hyperspins are mapped onto binary Ising spins right after a few initial time steps. In every run the nonlinear constant was fixed to $\beta=0.9$, while the equalization constant was set to $\delta_{eq}$=1 in the case of equalized runs and to zero otherwise. The results for the symmetry quenching scheme correspond to the energy readings obtained at the end of the first quenching cycle, while those from the Ising machine are read when the system reaches equilibrium (the absolute energy difference with the previous step is below $10^{-10}$ for $10$ time steps). The runs with quenching concentrate at small error regardless of equalization, with the equalized symmetry quenching set being the only one that reaches the ground state: $15$ of its $30$ runs indeed fall in the first bin (within 1\% of the benchmark ground state energy). The non-equalized quenching runs behave differently: all 30 of them ending on the same energy value that is above the benchmark minimum energy by a relative error of $1.1\times10^{-2}$. Instead, neither Ising machine setup reaches a comparable performance: over the $60$ runs the best relative errors are $1.1\times10^{-2}$ and $1.1\times10^{-1}$, respectively without and with equalization, with medians of $4\times10^{-2}$ and $3\times10^{-1}$.}
    \label{fig:Hist_Ann_vs_Cim_g05_600}
\end{figure}

Even if for some particularly hard instances of the considered graph set the machine does not provide a good SP with the protocol we tested, the symmetry quenching method still proves to greatly enhance the machine performance. We compare the energy readings at the end of the first quenching cycle with the results obtained using the machine as a bare Ising machine, keeping a fixed pump above threshold that immediately damps the imaginary components, and only reading the energy when the system reaches equilibrium using the condition in Eq.~\eqref{eqn:Energy_Difference}. In Fig.~\ref{fig:Hist_Ann_vs_Cim_g05_600} one can indeed see the difference between the energy error relative to the known ground-state value [Eq.~\eqref{eqn:Max-Cut_to_Ising}] across several runs, in the case of both the equalized and non-equalized quenching protocols (green and orange bars), and the equalized and non-equalized Ising machine (blue and pink bars), on a graph instance for which convergence to low-energy solutions was particularly hard. Specifically, we observe that, among the compared protocols, the only one that is able to reach the known ground state is the equalized quenching protocol.
This result also emphasizes that equalization is of great importance in order for the machine to properly complete minimization tasks.

\section{Scaling to one million hyperspins}\label{sec:Discussion}

Having assessed the capability of the complex hyperspin machine to find satisfying results in minimizing the complex $D$-vector Hamiltonian in Eq.~\eqref{eqn:D_Vector_Hamiltonian} in previous sections, we now show that the simulated machine is capable of handling systems of large sizes, up to $10^6$ spins. In this section we do not benchmark the quality of the solution reached by the complex hyperspin machine, since a meaningful performance assessment requires running other minimization algorithms (e.g., the DE numerical method used before) to find $H_{\rm min}$ on such large systems, which is not feasible with our numerical resources. The value of $H_{\rm min}$ is then defined as the minimum energy value found by the hyperspin machine out of the several simulations of the dynamics (both equalized and non-equalized) that we perform to collect data for the statistical analysis.

Additionally, the main obstacle when simulating large systems with dense matrices comes both from the huge memory requirements to store the coupling matrix and from increased simulation times. For this reason, we limited this first test at $N=10^6$ two-dimensional ($D$=2) complex hyperspins to sparse square lattices, which have a fixed connectivity (four edges on each node) and allow us to store the coupling matrix as sparse by reducing memory requirements. We generate a Hermitian lattice using $\alpha=0.50$ and use it to perform several dynamics runs. The parameters of the runs were fixed except for the equalization constant: $10$ runs were performed with $\delta_{eq}=0$ and $25$ with $\delta_{eq}=1$.

\begin{figure} 
    \centering
    \includegraphics[width=14cm]{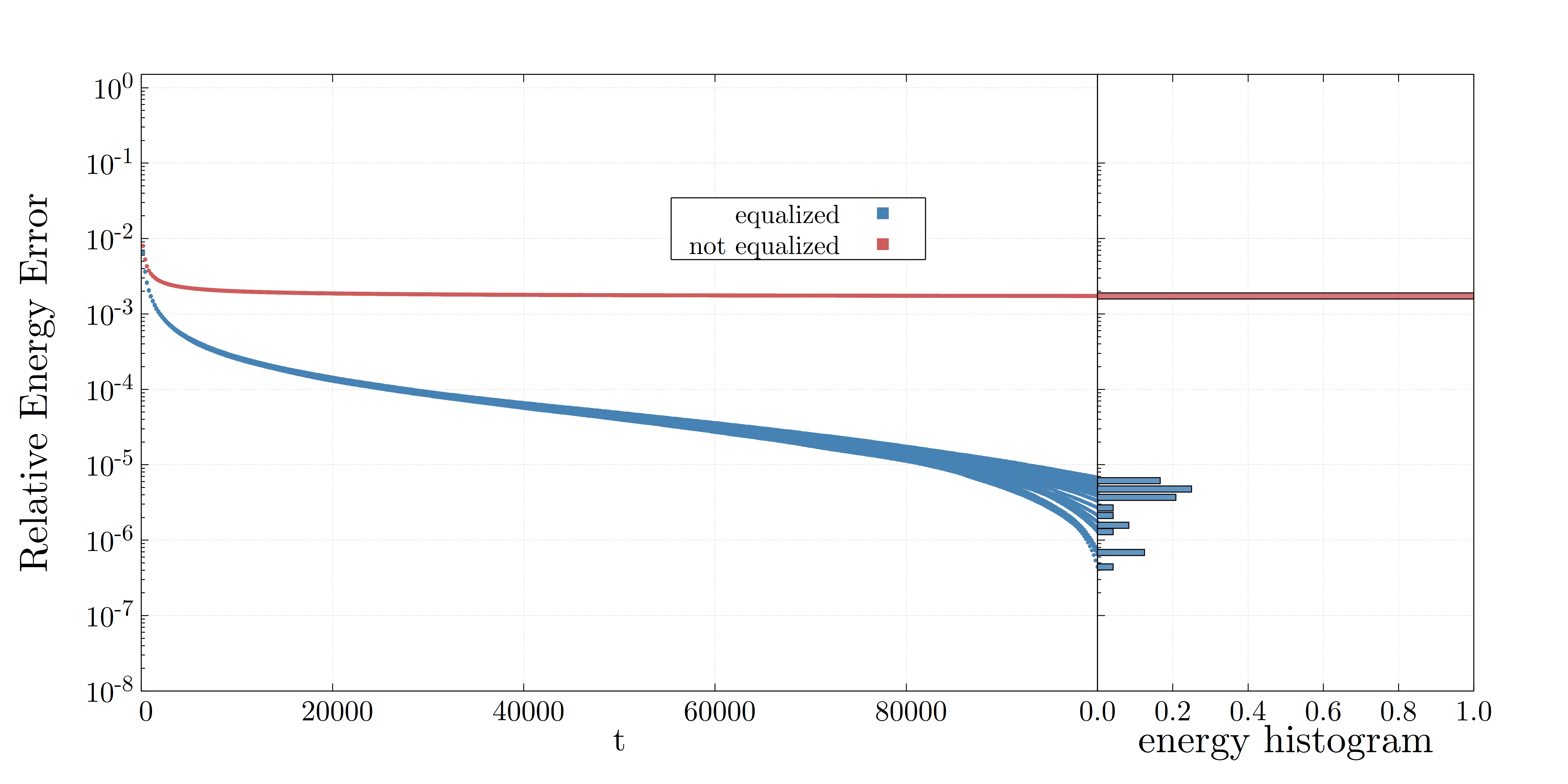}
    \caption{Relative energy error in Eq.~\eqref{eqn:Relative_Error} shown in log-linear scale for a system of $N=10^{6}$, $D=2$ complex hyperspins on a single sparse two-dimensional periodic square lattice, with Hermitian couplings generated at $\alpha=0.50$. The system is simulated in the harmonic regime ($h=0$) with no intrinsic loss ($g=0$) and using a nonlinear coefficient $\beta=0.1$. Integration is performed with time step $dt=0.1$, starting from $t=0$ up to $t=10^5$ ($10^{6}$ integration steps). A total of $35$ runs are performed, including both without equalization (red) and with equalization using $\delta_{eq}=1$ (blue). The relative energy deviation is computed using as a reference the lowest found energy $H_{\rm min}$ among all runs. (Left) Time evolution of the relative energy deviation. (Right) Histogram of the final energy values from both non-equalized (red bars) and equalized (blue) simulations. The equalized runs consistently reach lower energies with relative differences with the best found energy below $10^{-6}$, while non-equalized runs all converge around a relative difference of $10^{-3}$. The larger spread in the case of the equalized runs is an artefact of the logarithmic scale. While the non-equalized runs all converge at a comparable higher energy, the equalized runs have differences between order of $10^{-5}$ and $10^{-7}$. This is due to the fact that the system is still evolving at $t=10^{5}$ and would require larger simulation time to fully converge to a steady state within numerical precision. The short simulation time is however chosen in order to favor larger statistics at the expense of a better settling, to demonstrate the capability of the machine to handle very large systems ($8\times 10^{6}$ total dynamical variables). Simulations are run over $16$ threads on an 11th-generation Intel Core i9-11900K (nominal $3.50$ GHz) and required approximately $3.76\times 10^{4}$ s of wall-clock time (approximately $5.86\times10^{5}$ s of CPU time).}
    \label{fig:1M_Run_Dyn_Hist}
\end{figure}

Figure~\ref{fig:1M_Run_Dyn_Hist} shows the numerical results of the relative energy error, comparing equalized and non-equalized runs, and related energy histogram. The equalized runs consistently reach an energy value of $H\sim-7.809\times10^{5}$ whereas the non-equalized ones all stall around $H\sim-7.795\times10^{5}$. This leads to an average energy difference that spans between two and four orders of magnitude in log scale, again proving that equalization plays a fundamental role to converge to low-energy values. We observe that the relative energy error from the non-equalized machine runs stays above $10^{-3}$, while data from the equalized runs all sit between $10^{-5}$ and $10^{-6}$. In order to favor larger statistics we used a rather short total simulation time, only reaching $t=10^{5}$, for which the system dynamics has not converged yet to a steady state. Simulations are run on an 11th-generation Intel Core i9-11900K processor (nominal 3.50 GHz) using 16 parallel threads. With this setup each equalized run required a total wall-clock time of about $3.76\times 10^{4}$ s, corresponding to a total CPU time of $5.86\times10^{5}$ seconds.

\section{Conclusion}\label{sec:Conclusion}

In this work, we introduced the equalized complex hyperspin machine, by expanding the framework of multidimensional spin simulators to accommodate both complex fields with general Hermitian interaction graphs and amplitude equalization feedback. Specifically, we showed that the machine is able to naturally deal with optimization problems that can be cast in the form of a $D$-vector model Hamiltonian with complex valued spins. A dedicated C-language software development kit was built to simulate the network of oscillators and assess the machine performance.

We discussed the different behavior of the complex system with and without parametric drive, two regimes that we named harmonic and parametric phases, respectively. In the harmonic regime, the system exhibits a $U(1)$ symmetry, which reduces to $\mathbb{Z}_2$ when the pump is turned on in the parametric regime. To operate the coupled oscillator network as a complex hyperspin machine to minimize vector spin models with general Hermitian coupling matrices, we restricted our analysis to the harmonic regime because we observed that the presence of a nonzero pump consistently drove the hyperspin energy to significantly higher values compared to the one obtained with zero pump. The complex hyperspin machine was shown to outperform another numerical minimization algorithm (differential evolution for fixed accuracy, precision, machine working precision, and number of search points) in terms of computational time and quality of the solution, where the test was performed over a set of custom generated Hermitian interaction matrices up to about 80 spins. 
 
When the couplings are instead real, we exploited the transition between harmonic and parametric regimes caused by the presence of the parametric amplification to introduce a new dimensionality reduction technique that we called symmetry quenching. The machine initially operates in the harmonic phase using the full complex nature of the oscillator fields when the pump is set to zero. By turning on the parametric pump, the system enters the parametric phase. In terms of hyperspin vectors, this abrupt change of symmetry, from $U(1)$ to $\mathbb{Z}_2$, corresponds to the dimensional reduction of a spin vector model from dimension $2D$ to $D$. The effectiveness of symmetry quenching was tested on Max-Cut problems with known solutions from the BiqMac library, casting them as Ising spin Hamiltonians. The obtained results were promising, showing that a symmetry quenching schedule outperforms the machine used with no symmetry quenching. 

Simulations on commodity hardware were pushed up to a system of $8\times 10^6$ degrees of freedom on a sparse graph with frustration in order to showcase the scalability of the algorithm. Our analysis showed that equalization greatly enhances the capability of the complex hyperspin machine to converge to low-energy values, reducing by orders of magnitude the relative energy difference compared to the lowest value determined by the non-equalized hyperspin machine. We could not compare the results with other minimization methods at such large scales with our resources due to the numerical complexity of the problem.

Our work places the equalized complex hyperspin machine as a competitive heuristics for the minimization of the $D$-vector model Hamiltonian with complex valued spins on large scales, allowing also the implementation of symmetry quenching protocols to enhance the performance of Ising machines.

Several future directions remain to be explored. First, it will be important to find systematic methods to determine optimal regimes involving all network's parameters to maximize the system performance in minimization tasks. Also, more benchmarks should be performed on different problems of increasing size and complexity, in order to further validate the machine's effectiveness. Another perspective is given by the possibility to add noise to the oscillator dynamics, in order to understand how the hyperspin machine in the stochastic regime can be used to study thermodynamical properties of multidimensional spin vector systems and phase transitions in general topology, which are relevant to the field of statistical physics. Additionally, it remains to be understood to what extent the complex hyperspin machine in the parametric regime (i.e., hyperspin machine with complex coupling matrices and nonzero pump) can be controlled to effectively yield a hyperspin state selector, in analogy to the recently proposed Ising selector machine in the case of Ising spins~\cite{Tosca_2026_Ising_Selector_Machine_by_Kerr_POs}. The experimental implementation of the hyperspin machine in photonic setups recently reported in Ref.~\cite{Veraldi_2026_hyperspin} opens the intriguing possibility to test the complex hyperspin machine using fully optical systems, to address both fundamental properties of spin systems in arbitrary dimension and for practical applications like phase retrieval~\cite{Wang_2026_Phase_Retrieval_via_Gain_Based_Photonic_XY_Hamilton_Optimization}.

\section*{Acknowledgements}
Funded by the European Union (HORIZON-ERC-2023-ADG HYPERSPIM project Grant No. 101139828). Views and opinions expressed are however those of the author(s) only and do not necessarily reflect those of the European Union or the European Research Council. Neither the European Union nor the granting authority can be held responsible for them.

\section*{Code and Data Availability}
The software development kit used to simulate the complex hyperspin machine is publicly available at \url{https://github.com/tommasopiacenti/Complex_Hyperspin_Machine}.

\printcredits

\bibliographystyle{unsrtnat}

\end{document}